# Nonlinear Gap Junctions Enable Long-Distance Propagation of Pulsating Calcium Waves in Astrocyte Networks

Mati Goldberg[1], Maurizio De Pittà[1], Vladislav Volman[2,3], Hugues Berry[4]*, Eshel Ben-Jacob[1,2]*

1 School of Physics and Astronomy, Tel Aviv University, Ramat Aviv, Israel,
2 Center for Theoretical Biological Physics, University of California San Diego, La Jolla, California, United States of America,
3 Computational Neurobiology Lab, The Salk Institute, La Jolla, California, United States of America,
4 Combining, INRIA Rhône-Alpes, Villeurbanne, France

* E-mail: eshel@tamar.tau.ac.il (EBJ); hugues.berry@inria.fr (HB)

**Abstract**

A new paradigm has recently emerged in brain science whereby communications between glial cells and neuron-glia interactions should be considered together with neurons and their networks to understand higher brain functions. In particular, astrocytes, the main type of glial cells in the cortex, have been shown to communicate with neurons and with each other. They are thought to form a gap-junction-coupled syncytium supporting cell-cell communication via propagating $Ca^{2+}$ waves. An identified mode of propagation is based on cytoplasm-to-cytoplasm transport of inositol trisphosphate ($IP_3$) through gap junctions that locally trigger $Ca^{2+}$ pulses via $IP_3$-dependent $Ca^{2+}$-induced $Ca^{2+}$ release. It is, however, currently unknown whether this intracellular route is able to support the propagation of long-distance regenerative $Ca^{2+}$ waves or is restricted to short-distance signaling. Furthermore, the influence of the intracellular signaling dynamics on intercellular propagation remains to be understood. In this work, we propose a model of the gap-junctional route for intercellular $Ca^{2+}$ wave propagation in astrocytes. Our model yields two major predictions. First, we show that long-distance regenerative signaling requires nonlinear coupling in the gap junctions. Second, we show that even with nonlinear gap junctions, long-distance regenerative signaling is favored when the internal $Ca^{2+}$ dynamics implements frequency modulation-encoding oscillations with pulsating dynamics, while amplitude modulation-encoding dynamics tends to restrict the propagation range. As a result, spatially heterogeneous molecular properties and/or weak couplings are shown to give rise to rich spatiotemporal dynamics that support complex propagation behaviors. These results shed new light on the mechanisms implicated in the propagation of $Ca^{2+}$ waves across astrocytes and precise the conditions under which glial cells may participate in information processing in the brain.

## Author Summary

In recent years, the focus of Cellular Neuroscience has progressively stopped only being on neurons but started to include glial cells as well. Indeed, astrocytes, the main type of glial cells in the cortex, dynamically modulate neuron excitability and control the flow of information across synapses. Moreover, astrocytes have been shown to communicate with each other over long distances using calcium waves. These waves spread from cell to cell via molecular gates called gap junctions, which connect neighboring astrocytes. In this work, we used a computer model to question what biophysical mechanisms could support long-distance propagation of $Ca^{2+}$ wave signaling. The model shows that the coupling function of the gap junction must be non-linear and include a threshold. This prediction is largely unexpected, as gap junctions are classically considered to implement linear functions. Recent experimental observations, however, suggest their operation could actually be more complex, in agreement with our prediction. The model also shows that the distance traveled by waves depends on characteristics of the internal astrocyte dynamics. In particular, long-distance propagation is facilitated when internal calcium oscillations are in their frequency-modulation encoding mode and are pulsating. Hence, this work provides testable experimental predictions to decipher long-distance communication between astrocytes.

# Introduction

The 20th century witnessed crystallization of the neuronal doctrine, viewing neuron as the fundamental building block responsible for higher brain functions. Yet, neurons are not the only cells in the brain. In fact, almost 50% of the cells in the human brain are glial cells [1,2]. Due to their apparent lack of fast electrical excitability, the potential importance of glial cells in neural computation was downgraded in favor of the critical role played by these cells in neural metabolism. Recent experimental evidence however suggests that glial cells provide a role much more than support, including control of synapse function and formation, adult neurogenesis and regulation of cerebral blood flow (see e.g. [3] for a review). As a consequence, a new paradigm is emerging in brain science, according to which glial cells should be considered on a par with neurons.

In particular, astrocytes, the main type of glial cells in the cortex, have attracted much attention because they have been shown to communicate with neurons and with each other. Indeed, astrocytes can integrate neuronal inputs and modulate the synaptic activity between two neurons [4]. Neurotransmitters released from pre-synaptic neurons can bind to specific receptors on the astrocyte membrane and evoke $Ca^{2+}$ elevations in the astrocyte cytoplasm [5]. In turn, these activated astrocytes may release gliotransmitters, including glutamate and ATP, which feed back onto the synaptic terminals and modulate neuron responses [6].

Two main types of neuronal activity-dependent $Ca^{2+}$ responses are observed in astrocytes [7,8]: (1) transient $Ca^{2+}$ increases that are restricted to the very extremity of their distal processes [9,10] and (2) $Ca^{2+}$ elevations propagating along these processes as regenerative $Ca^{2+}$ waves, eventually reaching the cell soma. The latter kind of event can even propagate to neighboring astrocytes, thus forming intercellular $Ca^{2+}$ waves [11,12]. Although intercellular $Ca^{2+}$ waves have been extensively observed in astrocyte cultures [13,14], recent experimental evidence supports the possibility that they could also occur under physiological conditions [6], with propagation distances ranging from four [15] to up to 30 astrocytes [16]. These results therefore indicate that waves in astrocytes may represent an effective form of intercellular signaling in the central nervous system [17,18]. But further, they almost irresistibly bring about the hypothesis that this persistent astrocyte wave-based signaling could extend the repertoire of neural network communications, adding non-local interactions, both in space and in time [3].

In order to assess this hypothesis though, several aspects of $Ca^{2+}$ signaling in astrocytes remain to be elucidated. Experimental data suggest that a stimulus impinging on an astrocyte is preferentially encoded in the modulation of the frequency (FM) of astrocytic $Ca^{2+}$ oscillations [10]. This type of oscillations is often characterized by pulsating waves, i.e. the propagation of peak waveforms, with width smaller than period. However, the possibility of amplitude modulation (AM) or even coexisting AM and FM (AFM) encoding have also been inferred [19,20]. Actually, the frequency and amplitude of astrocytic $Ca^{2+}$ oscillations can be highly variable, depending on cell-specific properties such as $Ca^{2+}$ content of the intracellular stores, or the spatial distribution, density and activity of (sarco-)endoplasmic reticulum $Ca^{2+}$-ATPase (SERCA) pumps [21,22]. Yet, the propagation of wave-like signalling in the context of such great variability is yet not fully understood [14].

Much effort has also been devoted to understand the mechanisms responsible for initiation and propagation of intercellular $Ca^{2+}$ waves. From a single-cell point of view, intracellular $Ca^{2+}$ dynamics in astrocytes is mainly due to $Ca^{2+}$-induced $Ca^{2+}$ release (CICR) from the endoplasmic reticulum (ER) stores and its regulation by inositol trisphosphate ($IP_3$) [6]. But for the transmission of these internal signals from one astrocyte to the other, two possible routes have been uncovered. The first one involves the transfer of $IP_3$ molecules directly from the cytosol of an astrocyte to that of an adjacent one through gap junction intercellular hemichannels [23]. In the second route instead, propagation is mediated by extracellular diffusion of ATP which binds to plasma membrane receptors on neighboring astrocytes and regulates $IP_3$ levels therein [18,24]. Although these two routes need not be mutually exclusive, experiments indicated that intracellular propagation through gap junctions is likely the predominant signaling route in many astrocyte types [19,25-27].

Albeit experimental protocols monitor wave propagation as variations of intracellular $Ca^{2+}$, the molecule that is transmitted through gap junctions to neighboring astrocytes is not $Ca^{2+}$, but $IP_3$ [27]. Indeed, when the $IP_3$ in a given cell increases, some of it can be transported through a gap junction to a neighbor astrocyte. This $IP_3$ surge in the neighbor cell can in turn trigger CICR, thus regenerating the original $Ca^{2+}$ signal. Yet, the transported $IP_3$ is required to reach a minimal threshold concentration to trigger CICR in the neighboring cell. If this threshold is not reached, propagation ceases [28]. In this regard, previous theoretical studies stressed the importance of a mechanism for at least partial regeneration of $IP_3$ levels [29,30]. Such a mechanism, coupled with IP$_3$ transport, could induce local IP$_3$ concentrations large enough to trigger CICR [30], thus enabling Ca$^{2+}$ wave propagation. Production of IP$_3$ by Ca$^{2+}$-dependent PLCδ has been suggested as a plausible candidate regeneration mechanism [29,31,32]. However, the intercellular latencies of the Ca$^{2+}$ waves simulated with this mechanism are hardly reconcilable with experimental observations, hinting a critical role for gap junction IP$_3$ permeability [29,30].

In the present study, we investigated the intercellular propagation of Ca$^{2+}$ waves through the gap-junctional route by a computer model of one-dimensional astrocyte network. To account for intracellular Ca$^{2+}$ dynamics, we adopted the concise realistic description of IP$_3$-coupled Ca$^{2+}$ dynamics in astrocytes previously introduced in Ref. [33]. We specifically focused on the influence of gap junction linearity and internal Ca$^{2+}$ dynamics on the wave propagation distance. By means of bifurcation analysis and numerical solutions, we show that nonlinear coupling between astrocytes can indeed favor IP$_3$ partial regeneration thus promoting large-distance intercellular Ca$^{2+}$ wave propagation. Our study also shows that long-distance wave propagation critically depends on the nature of intracellular Ca$^{2+}$ encoding (i.e. whether Ca$^{2+}$ signals are FM or AM) and the spatial arrangement of the cells. Furthermore, our results suggest that, in the presence of weak coupling, nonlinear gap junctions could also explain the complex intracellular oscillation dynamics observed during intercellular Ca$^{2+}$ wave propagation in astrocyte networks [12].

## Methods

### The ChI model of intracellular Ca$^{2+}$ dynamics

We describe calcium dynamics in astrocytes by an extended version of the Li-Rinzel model [34], called the *ChI* model that we developed and studied in [33]. A detailed presentation of this model is also given in the Supplementary Information. Briefly, the *ChI* model accounts for the complex signaling pathway illustrated in Figure 1 that includes Ca$^{2+}$ regulation by IP$_3$-dependent CICR as well as IP$_3$ dynamics resulting from PLCδ-mediated synthesis and degradation by IP$_3$ 3-kinase (3K) and inositol polyphosphate (IP) 5-phosphatase (5P). The temporal evolution of astrocytic intracellular calcium in our model is described by three coupled nonlinear equations:

$$\frac{dC}{dt} = J_{chan}(C, h, IP_3) + J_{leak}(C) - J_{pump}(C) \tag{1}$$

$$\frac{dh}{dt} = \left(h_\infty(C, IP_3) - h\right)/\tau_h(C, IP_3) \tag{2}$$

$$\frac{dIP_3}{dt} = P_{PLC\delta}(C, IP_3) - D_{3K}(C, IP_3) - D_{5P}(IP_3) \tag{3}$$

in which the variables $C$, $h$, $IP_3$ represent the cell-averaged calcium concentration, the fraction of open IP$_3$R channels on the ER membrane, and the cell-averaged concentration of IP$_3$ second messenger, respectively. Each one of these variables is coupled to others via the set of equations that describe contributions of different biochemical pathways, as described in details in Supplementary Information (equations S1-S4) alongside the complete mathematical analysis of the model features.

In a single-cell context, this model reproduces most of the available experimental data related to calcium oscillations in astrocytes. In particular, it faithfully reproduces the experimentally reported changes of oscillation frequency and wave shape caused by SERCA pump activity modulations [22].

**Astrocyte coupling**

Experimental evidence shows that chemical signaling between astrocytes usually takes the form of propagating Ca$^{2+}$ pulses that are elicited following the gap-junctional transfer of IP$_3$ second messenger molecules [13]. Intracellular IP$_3$ activates the CICR pathway, giving rise to the observed rapid transient elevations in cytosolic free calcium. We considered three scenarios to describe the exchange of IP$_3$ between a pair of adjacent astrocytes: (1) linear, (2) threshold-linear (composed of a linear term operating after a threshold) and (3) non-linear (here described as sigmoid) coupling (see Figure 2). The linear model is a simple diffusive coupling; however, threshold-linear and non-linear models both transfer IP$_3$ only when the IP$_3$ gradient between the two adjacent cells overcomes a threshold value.

Our investigation of nonlinear coupling case was motivated by the experimental observations suggesting that gap junction permeability in itself can be actively modulated by various factors, among them different second messengers. Indeed, there is growing evidence that gap junctions may have greater selectivity and more active gating properties than previously recognized [35]. Several signaling pathways are able to modulate junctional permeability. In particular, the conductance state of Cx43, the main type of connexin in astrocyte gap junctions [36] is regulated by phosphorylation by PKC, which is also involved in IP$_3$ degradation [37,38], as well as by intracellular Ca$^{2+}$ [39]. These data suggest that astrocyte gap junction gating could be coupled to intra- and inter-cellular IP$_3$ and Ca$^{2+}$ dynamics [40,41] in a nontrivial fashion. Accordingly, several previous simulation studies have explored the influence of complex (e.g. regulated by second messengers) gap junctions [42-44]. We explore here their effects on intracellular Ca$^{2+}$ wave propagation in astrocytes.

**Linear coupling function**. The linear model simply results from Fick's law of diffusion. The flux $J_{i \rightarrow j}$ of IP$_3$ molecules (where $i, j$ are indices of adjacent model astrocytes) is proportional to their concentration gradient:

$$J_{i \rightarrow j} = F \cdot \Delta_{ij} IP_3 \tag{4}$$

where $\Delta_{ij} IP_3 = IP_3^i - IP_3^j$. Such coupling function is the standard model for a gap junction acting as a passive channel [30]. The coupling strength (or permeability) $F$ depends on the number of gap junction channels and their unitary permeability, and in what follows, it will be considered as a parameter.

**Non-linear coupling functions: sigmoid and threshold-linear.** Threshold-linear coupling only partially keeps the linear characteristics of the "classical" gap junction adding a threshold on IP$_3$ gradient below which the flux $J_{i \rightarrow j}$ is zero. On the other hand, sigmoid coupling adds a further saturating threshold on the IP$_3$ gradient value, above which the IP$_3$ flux is constant.

Sigmoid coupling is defined as:

$$J_{i \rightarrow j} = \frac{F}{2}\left(1 + \tanh\left(\frac{|\Delta_{ij} IP_3| - IP_3^{thr}}{IP_3^{scale}}\right)\right) \frac{\Delta_{ij} IP_3}{|\Delta_{ij} IP_3|} \tag{5}$$

where $F$ is the coupling factor, $IP_3^{thr}$ is the predetermined threshold value and $IP_3^{scale}$ is the width of the transition zone in the sigmoid function (see Figure 2). In order to allow comparison of the effects of the threshold-linear coupling function with that of the sigmoid one, the slope of the threshold-linear function was chosen to be coincident with that of the sigmoid coupling function (Figure 2). Threshold linear coupling is thus defined as:

$$J_{i \to j} = \frac{F}{2} \cdot \Theta\left(\frac{\left|\Delta_{ij} IP_3\right| - IP_3^{thr} - IP_3^{scale}}{IP_3^{scale}}\right) \frac{\Delta_{ij} IP_3}{\left|\Delta_{ij} IP_3\right|} \quad (6)$$

with the threshold function $\Theta(x) = x$ if $x > 0$, and 0 otherwise.

**The network model**

We consider chains of $N$ astrocytes where each astrocyte is coupled to its two nearest neighbors via gap junctions. Each $i$-th astrocyte ($i = 1,\ldots,N$) is associated with three variables $C^i$, $h^i$ and $IP_3^i$, that are respectively the cytosolic $Ca^{2+}$ concentration, the ratio of open IP$_3$Rs and the intracellular IP$_3$ concentration in this astrocyte. The dynamics of these internal variables is given by the *ChI* model (equations 1-3 and Supplementary Information for a detailed explanation):

$\forall i = 1,\ldots,N$,

$$\frac{dC^i}{dt} = J_{chan}(C^i, h^i, IP_3^i) + J_{leak}(C^i) - J_{pump}(C^i) \quad (7)$$

$$\frac{dh^i}{dt} = \left(h_\infty(C^i, IP_3^i) - h^i\right)/\tau_h(C^i, IP_3^i) \quad (8)$$

For all cells that are not at the boundaries of the astrocyte chain (i.e. $\forall i = 2,\ldots,N-1$):

$$\frac{dIP_3^i}{dt} = \left.\frac{dIP_3^i}{dt}\right|_r + J_{i-1 \to i} + J_{i+1 \to i} \quad (9)$$

where the internal reaction term for $IP_3$, i.e. $\left.\frac{dIP_3^i}{dt}\right|_r$, is given by equation (3). By contrast, the equations for the first and last cells, namely cell 1 and $N$, depend on the boundary conditions. We considered three types of boundary conditions: (1) reflective, (2) absorbing and (3) periodic. Reflective (zero-flux) boundaries assume that IP$_3$ exiting cell 1 or $N$ can only flow to cell 2 or $N$-1, respectively. They are given by:

$$\begin{cases} \frac{dIP_3^1}{dt} = \left.\frac{dIP_3^1}{dt}\right|_r + J_{2 \to 1} \\ \frac{dIP_3^N}{dt} = \left.\frac{dIP_3^N}{dt}\right|_r + J_{N-1 \to N} \end{cases} \quad (10)$$

We also considered the case where cells 1 and $N$ are absorbing, namely they entrap incoming IP$_3$ fluxes. This is the case of absorbing boundary conditions in which IP$_3$ can flow from cell 2 to cell 1, but the reverse flux (from cell 1 to 2) is always null (and similarly for cells N and N-1). Accordingly, equations read:

$$\begin{cases} \dfrac{dIP_3^1}{dt} = \dfrac{dIP_3^1}{dt}\bigg|_r + \Theta(J_{2\to1}) \\ \dfrac{dIP_3^2}{dt} = \dfrac{dIP_3^2}{dt}\bigg|_r + J_{3\to2} - \Theta(J_{2\to1}) \\ \dfrac{dIP_3^{N-1}}{dt} = \dfrac{dIP_3^{N-1}}{dt}\bigg|_r + J_{N-2\to N-1} - \Theta(J_{N-1\to N}) \\ \dfrac{dIP_3^N}{dt} = \dfrac{dIP_3^N}{dt}\bigg|_r + \Theta(J_{N-1\to N}) \end{cases} \quad (11)$$

Finally, with periodic boundary conditions, the 1D astrocyte chain actually takes the shape of a ring and the equations read:

$$\begin{cases} \dfrac{dIP_3^1}{dt} = \dfrac{dIP_3^1}{dt}\bigg|_r + J_{N\to1} + J_{2\to1} \\ \dfrac{dIP_3^N}{dt} = \dfrac{dIP_3^N}{dt}\bigg|_r + J_{N-1\to N} + J_{1\to N} \end{cases} \quad (12)$$

**Stimulation**

To induce wave propagation in the astrocyte chain, one cell (referred to as the "driving" cell) is stimulated by a supplementary exogenous IP$_3$ input. This external stimulus is supplied through a (virtual) "dummy" cell, coupled to the driving cell by one of the coupling functions described above. In this sense, the dummy cell acts as an IP$_3$ reservoir in which the level of IP$_3$ is kept fixed to a constant value $IP_3^{bias}$.

Let $k$ be the coordinate of the stimulated cell (driving cell) within the 1D chain. In this study, we usually stimulate the first cell or the central one, that is $k = 1$ or $k = (N+1)/2$ ($N$ odd). Hence, IP$_3$ dynamics in the $k$-th cell is given by

$$\dfrac{dIP_3^k}{dt} = \dfrac{dIP_3^k}{dt}\bigg|_r + J_{k-1\to k} + J_{k+1\to k} + J_{0\to k} \quad (13)$$

where $J_{0\to k}$ is calculated using $\Delta_{0k} IP_3 = IP_3^{bias} - IP_3^k$ in equation (4).

Most simulations done in this work were driven by a constant value of $IP_3^{bias}$. In the last section though, a square positive wave stimulus was applied to the model.

Initial conditions for all cells were set in agreement with experimental values reported in astrocytes for $Ca^{2+}$ and $IP_3$ at basal conditions [45].

**Numerics**

The chain model consists of $3N$ non-linear ODEs, where the number of astrocytes in the chain, $N$, ranged from 1 to 100. Time solutions were obtained via numerical integration by a standard 4$^{th}$-order Runge-Kutta scheme with a time step of 10 ms as this value showed to be the best compromise between integration time and robustness of the results. The computational model was implemented in Matlab (2009a, The MathWorks, Natick, MA) and C. Bifurcation analysis was done using XPPAUT (http://www.math.pitt.edu/~bard/xpp/xpp.html). Nonlinear time series analysis was performed using the TISEAN software package [46]. Table S1 in the Supplementary Information lists the values of the parameters used in the model.

**Results**

Before proceeding to study the propagation of calcium waves in spatially extended networks of astrocytes, it was necessary to understand the dynamical response of a single model cell in response to $IP_3$ stimulation. To this end, we performed a detailed bifurcation analysis of our model astrocytes. A wealth of dynamical regimes was discovered, allowing model astrocytes to encode information about $IP_3$ stimulus in amplitude-modulated (AM), frequency-modulated (FM) or mixed (AFM) modes, depending on parameter values (see Text S1 in the Supplementary Information and refs. [47,48]). We then proceeded to study the bifurcation diagrams for systems of coupled model astrocytes (utilizing different types of coupling as detailed in Methods). Briefly, the bifurcation analysis showed the existence of FM pulse-like oscillatory regimes at low $IP_3^{bias}$ values, which can turn into complex oscillations for larger $IP_3^{bias}$ values. Because stable oscillation regimes could coexist in the bifurcation diagrams with stable fixed points, it could not be predicted from these diagrams whether an $IP_3$ input to the cell would trigger pulse-like oscillations or not, i.e. whether it would switch the system from the fixed point to the oscillatory regime. Thus, we resorted to extensive numerical simulations to investigate under what conditions one could observe propagation of $Ca^{2+}$ waves along the astrocyte chain.

In agreement with previous studies (see [33] for a review), $IP_3$-triggered CICR indeed allows intercellular $Ca^{2+}$ wave propagation in our modeling framework, as shown in Supplementary

Information Figure S3. However, the range of wave propagation was usually restricted to and depended on the biophysical parameters that determine the profile of intracellular IP$_3$ dynamics [32]. In what follows, we delineate the role that these parameters play in Ca$^{2+}$ wave propagation, under different coupling modes (linear vs. nonlinear) and different encoding regimes (FM vs. AFM).

**Calcium wave propagation along astrocyte chains**

**Linear vs. non-linear gap junctions.** Propagation patterns both for the linear and nonlinear cases are presented in Figure 3 for the case of FM-encoding cells ($N = 12$). Here, a constant stimulus ($IP_3^{bias} = 1.0$ μM) was used and always applied to the first cell of the chain. Model analysis (see Methods) predicted that this level of IP$_3$ would trigger periodic Ca$^{2+}$ pulses at least in the stimulated cell and possibly in the other ones as indeed confirmed by simulations (see Astrocyte 1) both in the case of linear and non-linear coupling. For the linear coupling case, we observed propagation failure at 6$^{th}$-7$^{th}$ cell from the driving one (Figures 3a,b). By contrast, in the nonlinear coupling scenario, Ca$^{2+}$ pulses can propagate for the whole length of the chain (Figures 3c,d).

Analysis of the IP$_3$ pattern for the nonlinear coupling function (Figure 3d) evidences a strong correlation between IP$_3$ and Ca$^{2+}$ pulses. The IP$_3$ pulses are followed in time by the Ca$^{2+}$ ones, suggesting that pulsed Ca$^{2+}$ propagation is mediated by the propagation of IP$_3$ across the cells. By contrast, in the case of linear coupling, the correlation between the propagating Ca$^{2+}$ pulses and the intracellular IP$_3$ signals is not so apparent (Figures 3a,b) as IP$_3$ seems to diffuse smoothly from the stimulated cell without any effective propagation pattern.

The observed difference in the propagation distance between linear and nonlinear gap-junction couplings can be understood from this analysis. Indeed, in the case of linear gap-junction coupled cells, the IP$_3$ arriving in cell *i* from cell *i*-1 is transferred forward to cell *i*+1 before it can significantly accumulate in cell *i*. As a result, the IP$_3$ displays the almost diffusive pattern of Figure 3b, with a fast decay as the distance from the stimulated cell increases, and not real travelling wave structure in space. Hence even with large values of the coupling strength or stimulus intensity, beyond a limited number of cells away from the stimulated one, the IP$_3$ concentration becomes too small to trigger CICR. This stops Ca$^{2+}$ wave propagation. Conversely, with nonlinear gap junctions, IP$_3$ can accumulate in cell *i* (and trigger CICR) before it reaches the gap junction threshold and gets transferred to cell $i+1$. As a result, the IP$_3$ concentration evolves to the locally regenerative spatiotemporal pattern illustrated in Figure 3d) that allows Ca$^{2+}$ wave propagation over the whole network.

The distances (measured in units of number of cells) travelled by the propagating Ca$^{2+}$ waves as a function of the stimulation amplitude for an astrocyte chain of $N = 25$ FM-encoding cells are reported in Figure 4a. For linear gap junctions, the propagation distance increases with $IP_3^{bias}$, but never exceeds one third of the chain length. On the contrary, with nonlinear sigmoid coupling, Ca$^{2+}$ oscillations propagate along the whole chain as soon as the oscillatory regime is engaged (that is for $IP_3^{bias} > 0.72$ µM, see Figure S2).

Figure 4a further shows that threshold-linear gap junctions exhibit almost the same response as sigmoid ones but with different effective threshold IP$_3$ concentrations. Hence, these results indicate that the significant parameter for long-distance wave propagation through nonlinear gap junctions is the presence of an IP$_3$ concentration threshold below which the junction is closed (this property is shared by the two nonlinear models), rather than the saturation of the transport at high IP$_3$ concentrations (found only in the sigmoid model).

Because the effects due to the different shapes of coupling curves could be conflated in the above observations, we computed the dependence of wave propagation range on the maximal strength of coupling, for linear vs. nonlinear coupling cases (Figure 4c). For the most part of the range of examined coupling strengths, Ca$^{2+}$ wave propagation distance was significantly larger for the nonlinear case as compared to the linear case, ruling out the possibility that our findings are just a trivial confound. The only exceptions to this claim were noted for low $F$ values (these dynamics at low coupling are studied thereafter) and for very small regions (around $F = 1.5$ and $F = 2.5$), were the propagation distances for both couplings were comparable. More importantly, Figure 4c evidences that linear gap junction fails to propagate long-distance Ca$^{2+}$ waves (except in the "chaotic" low $F$ domain).

Figure 4d illustrates propagation ranges in for FM cells when $v_\delta$ (max. rate of IP$_3$ production by PLCδ) and $r_{5P}$ (max. rate of IP$_3$ degradation by IP-5P) vary. We locate with black dots the ($v_\delta, r_{5P}$) pairs for which Ca$^{2+}$ waves propagate across the whole cell chain. Clearly, long-range propagation is found for a wide region of this parameter space. As expected, larger IP$_3$ synthesis rates must be balanced by larger IP$_3$ degradation rate to allow long range propagation, hence the diagonal-like aspect of the black region in the panel.

These first results thus indicate that the propagation distance of $Ca^{2+}$ waves in our model is much smaller with linear gap junctions than with nonlinear ones. This observation remains valid when the number of cells in the chain is much larger (we have simulated up to 120 cells in the chains) or/and when up to the 20 first cells in the chain receive the stimulation simultaneously (not shown). The above results are also robust with respect to the changes in boundary conditions (see Methods). For instance, Figure S4 illustrates long-distance $Ca^{2+}$ wave propagation for a chain of $N = 12$ FM-encoding astrocytes with periodic or absorbing boundary conditions and gap junctions endowed with sigmoid-like coupling.

This confirms that the difference of propagation distance between linear and nonlinear gap junction-coupling is a robust and fundamental property of our model. Hence, the existence of a threshold concentration for cell-to-cell $IP_3$ diffusion, similar to the one displayed by nonlinear gap junctions may be a critical factor for long-distance propagation of $Ca^{2+}$ waves across astrocytes. In what follows, we examine the influence of a second physiological characteristic of $Ca^{2+}$ signaling in astrocytes, namely their stimulus encoding mode (FM-encoding or AFM-encoding chains).

**AFM vs. FM cells.** As illustrated in Figure 5, $Ca^{2+}$ waves do not propagate in our model of AFM-type astrocyte chains. In Figure 5, large $Ca^{2+}$ variations are observed only in the driving cell (cell 1 in Figure 5), whereas the other astrocytes exhibit subthreshold $Ca^{2+}$ changes or no $Ca^{2+}$ change at all. Importantly, Figure 4b shows that this observation is not restricted to the parameters of Figure 5 but holds true whatever the stimulation strength of the driving cell or the nature of the connecting gap junctions are, i.e. linear or nonlinear. In particular Figure 4b reports propagation failure even when the stimulation applied to the driving cell is as strong as $IP_3^{bias} = 1.5$ μM, namely an intensity deeply inside the oscillatory region of the bifurcation diagrams in Figures S1a, c. This failure to propagate is not caused by a failure of $IP_3$ diffusion through gap junctions. First, in the case of nonlinear gap junctions, the stimulus strength is well beyond the diffusion threshold ($IP_3^{thr}$). Secondly, failure is also observed with linear gap junctions, where no coupling threshold can impede cell-to-cell diffusion. Hence the propagation failure likely stems from an intrinsic inability of AFM astrocytes to build up sufficient intracellular $IP_3$ levels to trigger CICR in neighboring cells.

This intrinsic difference in the propagation properties brought about by AFM or FM modes can be explained on the basis of the single-cell bifurcation diagrams (Figures S1a, c). Indeed, in the AFM-encoding mode, the peak concentration of the IP3 and $Ca^{2+}$ oscillations decreases with decreasing $IP_3$ stimulations. Hence the $IP_3$ generation in AFM is such that a local depression of transmitted $IP_3$ will be accentuated in the next cell. Any decline of $IP_3$ production in a given cell will thus be transmitted outward and amplified along the chain, until the signal eventually fails. This phenomenon is not observed with FM cells because, by definition of the FM mode, the peak amplitude of the $IP_3$ oscillations in cell $i$ is hardly dependent of the strength of the $IP_3$ stimulus coming from cell $i$-1 (at least in the limit where the incoming $IP_3$ stimulus falls within the oscillatory range of cell $i$). This simple mechanism guarantees that peak $IP_3$ values in cell $i$ will remain high even though the incoming stimulation is lower. In other words, the FM mode guarantees robust regeneration of the wave propagation.

Moreover, the range of $IP_3$ input that gives rise to oscillations in the AFM encoding regime is much narrower than in the FM case. Thus, a perturbation of the $IP_3$ stimulation from cell $i$ to $i$+1 in the AFM mode is more likely to be enough to push cell $i$+1 outside of its oscillatory range, leading to termination of wave propagation in this cell. Importantly, our simulations with AFM-encoding cell chains reported propagation failures for all the $F$ values (Figure 4c) and ($v_\delta, r_{5P}$) pairs that were tested (results not shown).

Therefore, these results suggest a neat functional difference between AFM and FM oscillations in astrocytes: while FM could support long distance propagation of pulse-like $Ca^{2+}$ waves, AFM is rather expected to give rise to localized $Ca^{2+}$ signalling with diffusion-like spatial patterns for $IP_3$. Hence, any parameter relevant to *intra*-cellular $Ca^{2+}$ signaling and able to switch the cell between AFM and FM modes (e.g. the affinity or activity of the SERCA pumps) is predicted to play a key role in the *inter*-cellular propagation of $Ca^{2+}$ signals in astrocytes.

**Propagation in composite chains.** Because the astrocyte population within the brain is heterogeneous [49], the results reported above question the possibility of intercellular $Ca^{2+}$ wave propagation across astrocytes with different properties. Here we tackled this issue using composite astrocyte chains, namely chains constituted of both FM and AFM cells, and investigated under what conditions propagation is possible with nonlinear sigmoid gap junctions.

In Figure 6, we stimulated the first cell of the chain (cell 1) with a constant stimulus so as to initiate the $Ca^{2+}$ wave in this cell. The intensity of the stimulus was set close to the upper edge of the cell oscillatory range according to the bifurcation diagrams in Figures S1c, d so as to maximize the chance of wave propagation. Figure 6a illustrates the propagation of the $Ca^{2+}$ wave in a chain of alternating FM (black traces) and AFM (gray traces) cells and shows that propagation abruptly terminates at the second AFM cell in the chain (cell 4). Notably, closer inspection of $IP_3$ dynamics in the subsequent FM cell (i.e. cell 5 in Figure 6d) reveals that the $IP_3$ concentration intermittently passes across the predicted threshold for oscillations in this cell. However the time spent above the threshold is never large enough to trigger CICR, so that propagation halts. Extending propagation to further cells in the chain thus demands faster endogenous $IP_3$ production. This can for instance be obtained with larger values of the maximal rate of PLCδ, $v_\delta$, and/or smaller $IP_3$ degradation rates, $v_{3K}$ and/or $r_{5P}$.

The former possibility is considered in Figures 6b,e. The simulation reported in these figures corresponds to the same conditions as in Figures 6a,d, except that $v_\delta$ is larger in AFM cells. Clearly, wave propagation now extends across the entire chain. Due to the increased rate of $IP_3$ production, all AFM cells in the chain in fact maintain intracellular $IP_3$ concentration either beyond or within the oscillatory range. Moreover, since this range essentially overlaps with the lower part of the oscillatory range for FM cells, the $IP_3$ transported from one AFM to the next FM cell in the chain can trigger CICR there, thus perpetuating propagation.

Another possible mechanism to facilitate wave propagation across AFM cells in heterogeneous conditions consists in increasing the frequency of the wave pulses in the FM cell preceding the AFM one. This effect is actually naturally obtained when several successive FM cells are placed between two AFM ones. We illustrate this in Figures 6c,f, where the same conditions as in Figures 6a,d were used, except that one has now two successive FM cells between two AFM ones. The interactions between the two successive FM cells increase the frequency of the $Ca^{2+}$ pulses, and thus the frequency of elementary diffusion events of $IP_3$ in the next AFM cell. In turn this increases both the frequency of the $IP_3$ oscillations in the AFM cell and their minimal level, thus allowing $Ca^{2+}$ wave propagation in the subsequent FM cells.

The presence of homogenous FM cell domains between AFM cells is therefore likely to enable long traveling distances for propagating $Ca^{2+}$ waves. One may even assume that if the number of successive FM cells in the FM domains is large enough, the $Ca^{2+}$ wave should propagate over the entire network, whatever its size. Although we did not further investigate this possibility, our simulations hint on the contrary that there likely exists an upper bound for the travelling distance because the frequency of propagating Ca$^{2+}$ waves in FM domains is not constant, but tend to decrease after each AFM cell, as can be seen by comparison of the pulse frequency in the Ca$^{2+}$ traces of FM cells in Figures 6a-c. Indeed such progressive decay of the pulse frequency along the chain eventually brings forth insufficient IP$_3$ diffusion through gap junctions, thus terminating propagation (results not shown).

We note that modifications of the IP$_3$ threshold for diffusion in the nonlinear gap junctions, $IP_3^{thr}$, should facilitate transmission of Ca$^{2+}$ pulses from cell to cell in the chain, thus increasing the frequency of propagation (Figure S5), with the possibility of observing very different dynamics of propagation in chains of identical astrocytes. Such scenario supports the notion that although nonlinear gap junctions could explain long-distance propagation, their specific properties are expected to be critical factors for the dynamics of propagation. This aspect is further investigated in the next and last section of results of our study.

**Propagation of complex waves**

The Ca$^{2+}$ and IP$_3$ dynamics observed so far were all obtained using a rather high value of the coupling strength ($F = 2.0$ μM·s$^{-1}$). In these conditions, the properties of the propagated waves are rather simple: a pulse-like (or not) wave front travels across astrocytes, with conserved shape and either stops after a few cells or invades the whole cell chain. However, our system is a spatially extended dynamical system with large numbers of degrees of freedom. Such systems (e.g. coupled map lattices) are known to manifest complex spatiotemporal behaviors when the coupling strength changes. To get an insight on the possible propagation behavior exhibited by our model with weaker coupling, we considered the dynamics with reduced levels of gap junction permeability (setting $F = 0.23$ μM·s$^{-1}$).

Figure 7 shows the Ca$^{2+}$ dynamics of 41 coupled FM cells and a square wave periodic stimulation applied to the central cell #21 (see figure caption for details). Visual inspection of the Ca$^{2+}$ traces in each cell (Figure 7a) indicates that such periodic (oscillatory) stimulation can trigger Ca$^{2+}$ waves that can propagate along the whole chain. Importantly, this figure also evidences the occurrence of occasional propagation failures that do not seem to result from a simple spatiotemporal pattern. Actually, observation of the temporal traces of each individual cell reveals the occurrence of pulse-like events showing up with no apparent regularity. Accordingly, the distribution of the time-intervals between two such pulses can be very broad for some cells, with large intervals often almost as probable as small ones (see Figure S6).

Albeit consistently pulse-like, the shape of the propagated $Ca^{2+}$ waves is also quite variable. Closer inspection of the time series for the driving cell (i.e. cell 21) for instance shows that the generated $Ca^{2+}$ pulses vary from a single-peak waveform to multiple peaks per single pulse (Figure 7c). Furthermore, Figure 7d shows that the variability and complexity of the $IP_3$ signals is also very large. The lack of obvious regular behavior is particularly striking on movies showing the parallel temporal evolution of the $Ca^{2+}$ and $IP_3$ level in each cell, as in Video S1 in the Supplementary Information.

To further illustrate the complexity of the obtained dynamics, we plot in Figure 7b the trajectory of the system in the phase space of the driving cell. It is very tempting to compare the resulting trajectories to those observed with classical low-dimensional strange attractors. In this regard, preliminary analysis of the three time series of the driving cell using nonlinear time series analysis tools [46] suggested that the dynamics indeed corresponds to deterministic chaos, with sensitivity to initial conditions testified by a positive maximal Lyapunov exponent that we estimated between 0.020 and 0.050 $s^{-1}$ (depending on the time series under consideration).

The apparent complexity of the dynamics is most likely due to some form of spatiotemporal chaos, the nature of which is beyond the scope of the current article and is left to future work. But whatever the response, these simulations evidence that complex $Ca^{2+}$ wave propagation patterns can manifest at low couplings, even with spatially homogeneous cell properties and in the absence of any stochasticity source.

## Discussion

Calcium-mediated signalling is a predominant mode of communication between astrocytes [50]. Consequently, it is important to understand how different biophysical mechanisms determine the ability of these brain cells to communicate over long distances. Here, we used the computational modeling approach to study the properties of gap junction-mediated signaling in simple networks of realistically modeled astrocytes. Using numerical simulations and tools of bifurcation theory, we showed that long-distance regenerative $Ca^{2+}$ wave propagation is possible when the gap junctions are rendered by nonlinear permeability but only when most of the model astrocytes are tuned to encode the strength of incoming IP$_3$ signal into frequency modulated Ca$^{2+}$ oscillations.

There has been a long-standing debate over the nature and characteristics of intercellular Ca$^{2+}$ waves observed in astrocyte networks. The present article concerns about the purely intracellular route, which involves the transfer of IP$_3$ molecules directly from cytosol to cytosol through gap junctions [23]. In the extracellular route instead, propagation is mediated by extracellular diffusion of ATP and purinergic receptor activation [18, 24]. Although these two routes need not be mutually exclusive, experiments indicated that their relative influences vary across the brain. Indeed, experimental evidence suggests that the purely intracellular route predominates in astrocytes of the neocortex [14,51] and the striatum [28] while the extracellular (purinergic-dependent) route seems predominant in the CA1 hippocampus area as well as in the corpus callosum [51]. Hence, the results obtained in the present paper are expected to be relevant to the former structures. Their relevance to the case where the two routes coexist could however be tested by simple extensions of our current model, in the spirit of the recent Ref. [52].

A critical issue for the modeling studies of intercellular Ca$^{2+}$ waves is to explain the observed variability of Ca$^{2+}$ wave travelling distance [14]. Indeed, experimental measurements show travelling distances varying from 30 cells [16] (that is, often outside the imaging microscope field) down to 3-4 cells only [15]. Models featuring purely regenerative waves (e.g. traveling waves in the usual mathematical sense) easily account for long distance propagations but hardly account for the observed short ones. Conversely, nonregenerative models (e.g. purely diffusive ones) cannot explain long-range propagation. A possible solution was suggested by Höfer et al. [32]. In the model proposed by these investigators, long-range propagating Ca$^{2+}$ waves are obtained via IP$_3$ regeneration in each cell by Ca$^{2+}$-activated PLC$\delta$. However, whenever PLC$\delta$ maximal activity is lower, regeneration becomes partial and the Ca$^{2+}$ wave propagation distance decreases. Yet this model does not include Ca$^{2+}$-dependent IP$_3$ degradation, which could be critical for the occurrence of IP$_3$-mediated Ca$^{2+}$ oscillations [33]. This latter process in particular, can compete with PLC$\delta$-mediated IP$_3$ production, thus hindering IP$_3$ regeneration and Ca$^{2+}$ wave propagation. This calls for additional factors to be taken into account to explain intercellular Ca$^{2+}$ wave propagation.

A first prediction of our model is that, regenerative waves are possible in a network composed in its majority of astrocytes that encode information about incoming IP$_3$ signals in the frequency of their Ca$^{2+}$ oscillations (FM). Interestingly, the response of astrocytes in vivo to IP$_3$ stimulation is known to exhibit high variability, both in frequency and amplitude [53-55]. This variability could be due to cell-to-cell heterogeneity (extrinsic noise) in some of the CICR parameters. In particular, this could include variability of the expression of PLCδ or of the affinity for Ca$^{2+}$ of the SERCA pumps. To our knowledge, the kinetic properties of SERCA2b have never been measured in astrocytes. However the hypothesis that SERCA2b affinity for Ca$^{2+}$ shows variability *in vivo* seems realistic, given the experimental literature. First, reports of experimental measurements of the SERCA2b affinity showed somewhat variable results, ranging from 170 [56] to 270 nM [57], albeit both studies used cDNA transfection in COS1 cells. Secondly, SERCA2b functionality can be directly modulated by quality-control chaperones of the ER, e.g. calreticulin and calnexin [58]. In particular, there exists strong indication that calreticulin may dynamically switch SERCA affinity for Ca$^{2+}$ from 170 to ~ 400 nM [59]. In this case, cell-to-cell variability in the concentration of calreticulin could result in the mixed AFM-FM cell networks studied here. Our observations then lead to the experimental prediction that such variability or heterogeneity of the astrocyte response would have a strong impact on the propagation of intercellular calcium waves between these cells. Notably, this scenario is also supported by several experimental studies [21,60]. In particular, calreticulin has been shown to regulate Ca$^{2+}$ wave propagation via direct interaction with SERCA2b thus modulation of Ca$^{2+}$ uptake by this pump [61].

In our model, the strength and the transfer properties of the gap junction coupling are critical permissive factors that allow long-range intercellular signaling between the astrocytes. In particular, nonlinear gap junctions were found to significantly enhance the range of Ca$^{2+}$ wave propagation (as opposed to the classic linear gap junctions that caused fast dissipation). Gap junctions with dynamic resistance are known to exist in cardiac networks [62,63] and in several other cells [35]. Yet there is currently no direct evidence for nonlinear transfer of second messenger molecules through gap junctions between astrocytes. Nonetheless, the activation of PKC, which is intimately related to IP$_3$ metabolism [33,64], is known to block astroglial gap junction communication and inhibit the spread of Ca$^{2+}$ waves therein [65]. Hence, in light of the existing knowledge regarding the control of gap junctional permeability by various signaling molecules [66], it is plausible to assume that some nonlinearity should exist in astrocytes too. The exact form of the nonlinearity of course will be dictated by the properties of the solute and the nature of its interaction with the membrane channels in the proximity of the gap junction complex. Meanwhile, the generic form of nonlinear coupling that we considered here allowed us to get a qualitative insight into the putative effect of nonlinear coupling on signal propagation in model astrocyte networks.

In the present study, we considered a simplified setup of 1D network implemented as a regular chain of coupled cells. Such 1D chains display attractive aspects. In particular, we could proceed to a numerical bifurcation study of these 1D coupled-cell systems (see Figure S2), which has proven invaluable for the interpretation of the simulation results. Such bifurcation analysis would hardly be possible in higher dimensions (e.g. 2D), because the number of cells one needs to account for in 2D is much larger than in 1D at constant propagation distance. Furthermore, a serious study of a 2D system must include the exploration of the influence of the coupling network topology [67], which adds further parameters to the study of the robustness of the model dynamical features. However, real astrocytes in tissues are believed to organize in quasi 2D networks with significantly more complex structure. Our model is thus a simplification of this quasi 2D reality. For instance, obstruction of wave propagation could dependent on the spatial dimension. Indeed in 2D or 3D reaction-diffusion systems or on random graphs, where the strength of the coupling or the local number of neighbors can vary across the network, the wave propagation distance can critically depend on the number of stimulated cells or the distribution of the number of coupled neighbors [68]. It is not yet clear whether our observation that linear gap junctions support only local wave propagation is restricted to regular 1D networks such as those used in the present work. Future works will be designed to tackle this issue. Nevertheless, in spite of its simplicity, this 1D model yields important predictions about the influence of the spatial arrangement of astrocytes. In particular, it shows that the distribution in space of heterogeneous gap junction permeabilities can result in rich dynamics [14,30]. Reducing the maximal strength of coupling between the model astrocytes imparted the individual cells with rich dynamics, possibly associated with spatiotemporal chaos. Keeping in mind that in reality the changes in gap junction permeability are mediated by the dynamic action of different effectors, we anticipate that a network of biological astrocytes could have the capacity to self-regulate the complexity of its dynamics. Whether or not this is the case, can be determined by experiments that selectively target the pathways of gap junction regulation.

Recent studies suggest that the astrocytes within the cortex form heterogeneous populations [69,70]. Therefore, we considered the case of intercellular $Ca^{2+}$ wave propagation in composite 1D networks, consisting of both FM- and AFM-encoding cells. Our simulations predict that the propagation dynamics and distance of intercellular $Ca^{2+}$ waves critically depends both on the encoding property of the cells and on their spatial arrangement. Interestingly, the cell bodies of neighboring astrocytes within the brain are believed to distribute in space in a nonrandom orderly fashion called "contact spacing" [71,72]. Our study thus suggests a possible link between contact spacing and intercellular $Ca^{2+}$ wave propagation in astrocyte networks. If, as suggested by our model, the spatial arrangement of the astrocytes, coupled to the heterogeneity of their response, conditions $Ca^{2+}$ wave propagation, then contact spacing may play a critical role in intercellular wave propagations in the brain and the related computational properties of astrocyte networks.

It is now widely accepted that astrocytes and neurons are interwoven into complex networks and are engaged in an intricate dialogue, exchanging information on molecular level [3]. By releasing different gliotransmitters (such as glutamate and ATP) astrocytes dynamically modulate the excitability of neurons and control the flow of information at synaptic terminals [4]. Diffusion of glutamate and/or ATP is limited due to the action of glutamate transporters and degradation of ATP, thus defining spatiotemporal range for the local effect of astrocyte on neurons and synapses [73]. On the other hand, long-range and temporally delayed regulation of neuronal and synaptic activity by astrocytes is believed to be mediated by intercellular $Ca^{2+}$ waves spreading through the astrocyte network. The connectivity of this astrocyte network is in turn defined by the patterns of electrical activity in neuronal network [74]. Thus, it appears that astrocytes and neurons are organized in networks that operate on distinct time scales and utilize the principles of feedback regulation to modulate the activities of each other. How such mutual regulation of neuronal and astrocytic networks affects the complexity of neuronal network dynamics in health and disease is a question that should be addressed by future combined experimental and modeling studies.

## Acknowledgments

The authors wish to thank Vladimir Parpura, Giorgio Carmignoto and Ilyia Bezprozvanny and Herbert Levine for insightful conversations.

**Figure 1**. **Sketch of the signaling pathways considered in the *ChI* model**. (**a**) $Ca^{2+}$-induced $Ca^{2+}$ release (CICR) from the endoplasmic reticulum (ER) is the main mechanism responsible for intracellular $Ca^{2+}$ dynamics in astrocytes. (**b**) Schematic of the coupling between $Ca^{2+}$ dynamics and $IP_3$ metabolism in the astrocyte. (**c**) Endogenous $IP_3$ production is brought forth by hydrolysis of $PIP_2$ by PLCδ (the activity of which is regulated by $Ca^{2+}$). (**d**) Degradation of $IP_3$ mainly occurs through $IP_3$ 3-kinase- (3K-) catalyzed phosphorylation and inositol polyphosphate 5-phosphatase (IP-5P)-mediated dephosphorylation. For simplicity, $Ca^{2+}$-dependent PKC-mediated phosphorylation of $IP_3$-3K [75] and competitive binding of $IP_4$ to IP-5P are not considered in this study. The legend of different arrows is given below (**c**).

**Figure 2**. **The coupling functions used in the current study to model different types of gap-junctions.** Shown is the relative flux, i.e. the value the $IP_3$ flux divided by the coupling force $F$ as a function of the $IP_3$ gradient ($\Delta IP_3$) between two coupled cells, for linear, threshold-linear and sigmoidal coupling. Parameters: $IP_3^{thr} = 0.3$ µM, $IP_3^{scale} = 0.05$ µM.

**Figure 3**. **Propagation patterns with linear (a, b) and non-linear sigmoidal (c, d) gap junctions.** The astrocyte chain was composed of 12 FM-encoding cells with reflective boundary conditions. Stimulation triggered by $IP_3^{bias} = 1.0$ µM from $t = 0$ s to $t = 120$ s applied to the first cell in the line.

**Figure 4**. **Traveled distance for the propagation of $Ca^{2+}$ waves as a function of stimulus intensity.** The stimulus is applied to the first cell and the traveled distance is expressed in number of cells. With moderate coupling strength ($F=2$ µm/s) and $N=25$ cells, long-range $Ca^{2+}$ propagation is observed in the case of FM (**a**) but not AFM chains (**b**). Linear gap junctions (*closed green circles*) do not sustain propagation over long distances, whatever the encoding mode is. Long-range propagation is observed for AFM cells coupled by non-linear sigmoid (*open magenta circles*) or threshold-linear (*times signs*) gap junctions. Variation of the propagation range with $F$ for 50 coupled FM cells ($IP_3^{bias} = 2.0$ µM) is shown in (**c**) for sigmoid and linear gap junctions. Panel (**d**) shows a map of the $r_{5P}$-$v_\delta$ parameter space, where black dots correspond to waves propagating over all the cells, while white areas locate non-propagating waves. In these figures, a wave was considered to have reached a given cell whenever the amplitude of calcium variations in this cell was larger than 0.6 µM. Boundary conditions in these simulations were reflective.

**Figure 5. Propagation patterns for linear (a, b) and non-linear sigmoidal (c, d) gap junctions.** The astrocyte chain was composed of 12 AFM astrocytes. Stimulus protocol and other parameters as in Figure 3.

**Figure 6. Calcium (a-c) and IP$_3$ (d-f) traces for wave propagation in composite astrocyte chains.** *Black* traces locate FM cells while AFM cells are displayed with *gray* traces. (**a**, **d**) Alternating one FM with one AFM cell for example, may not allow propagation beyond the second AFM cell. (**b**, **e**) Larger propagation distances can be obtained by increasing the maximal rate of PLCδ in AFM cells. (**c**, **f**) Alternatively, longer traveling distances can be observed in chains where two AFM cells are separated by more than one FM cell. Parameters: a stimulus of $IP_3^{bias}$ = 1.2 µM was applied between $t$ = 10 and $t$ = 300 s on the first cell of the chain (i.e. cell 1). Other parameters: $v_\delta^{(AFM)}$ = 0.108 µM/s; $v_\delta^{(FM)}$ = 0.832 µM/s; $r_{5P}$ = 0.202 s$^{-1}$; $IP_3^{thr}$ = 0.215 µM; (**b**, **e**): $v_\delta^{(AFM)}$ = 0.15 µM/s. In all these simulations, a 5-minute-long stimulus was applied.

**Figure 7. Complex behaviors at low coupling strength ($F$ = 0.23 µM·s$^{-1}$).** The stimulus is an oscillatory input (positive square wave) applied to the central cell of an $N$ = 41 cell chain. (**a**) Calcium concentration in cells 1 to 21 (cells 21 to 41, respectively, are identical). (**b**) The trajectory in the *C-h-IP$_3$* phase space for cell 21 (i.e. the stimulated cell) and (**c**, **d**) the corresponding Ca$^{2+}$ and IP$_3$ time series. Simulation performed on FM-encoding astrocytes with reflective boundary conditions and sigmoid gap junctions. Stimulus protocol: positive square wave of 50-second period and duty cycle of 0.4.

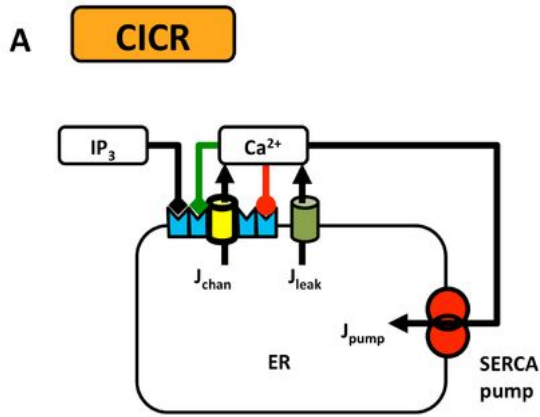
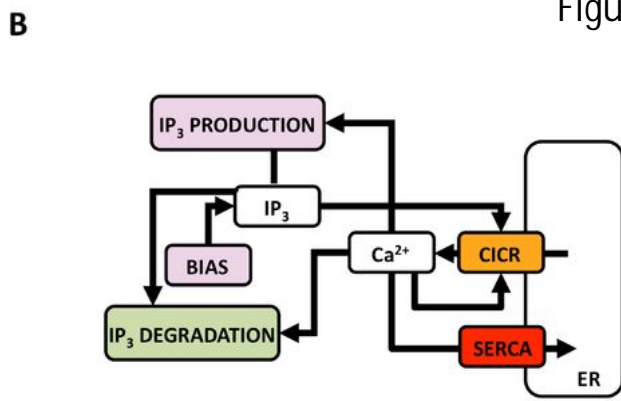
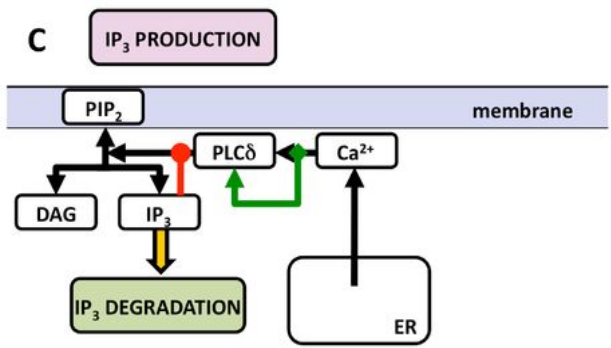
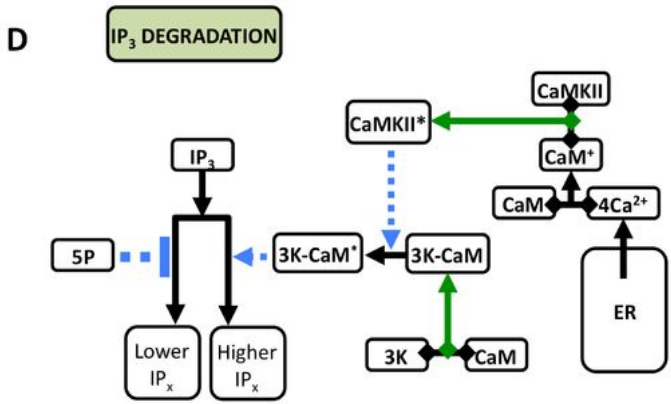
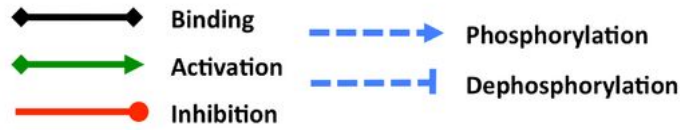

Figure 1

Figure 2

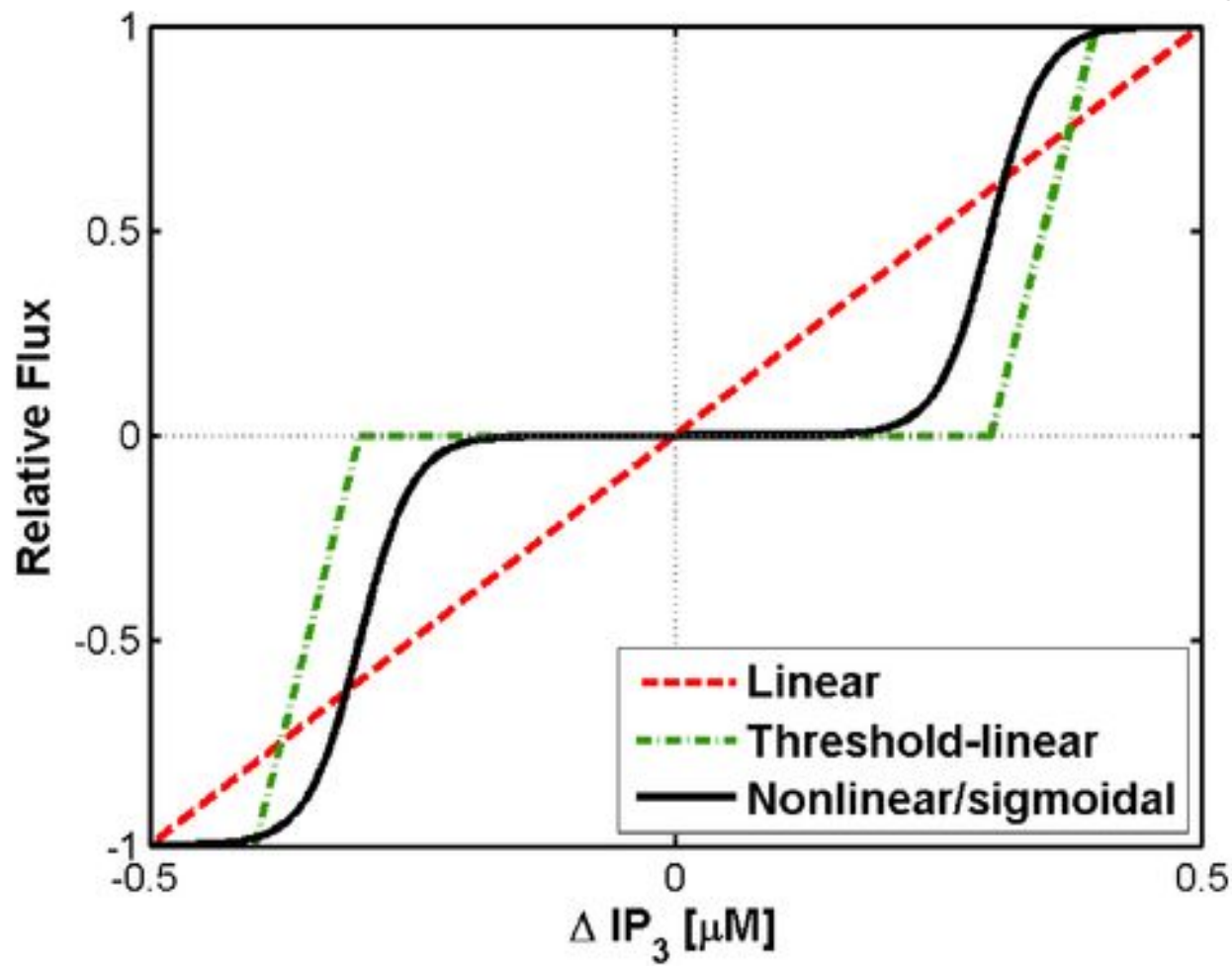

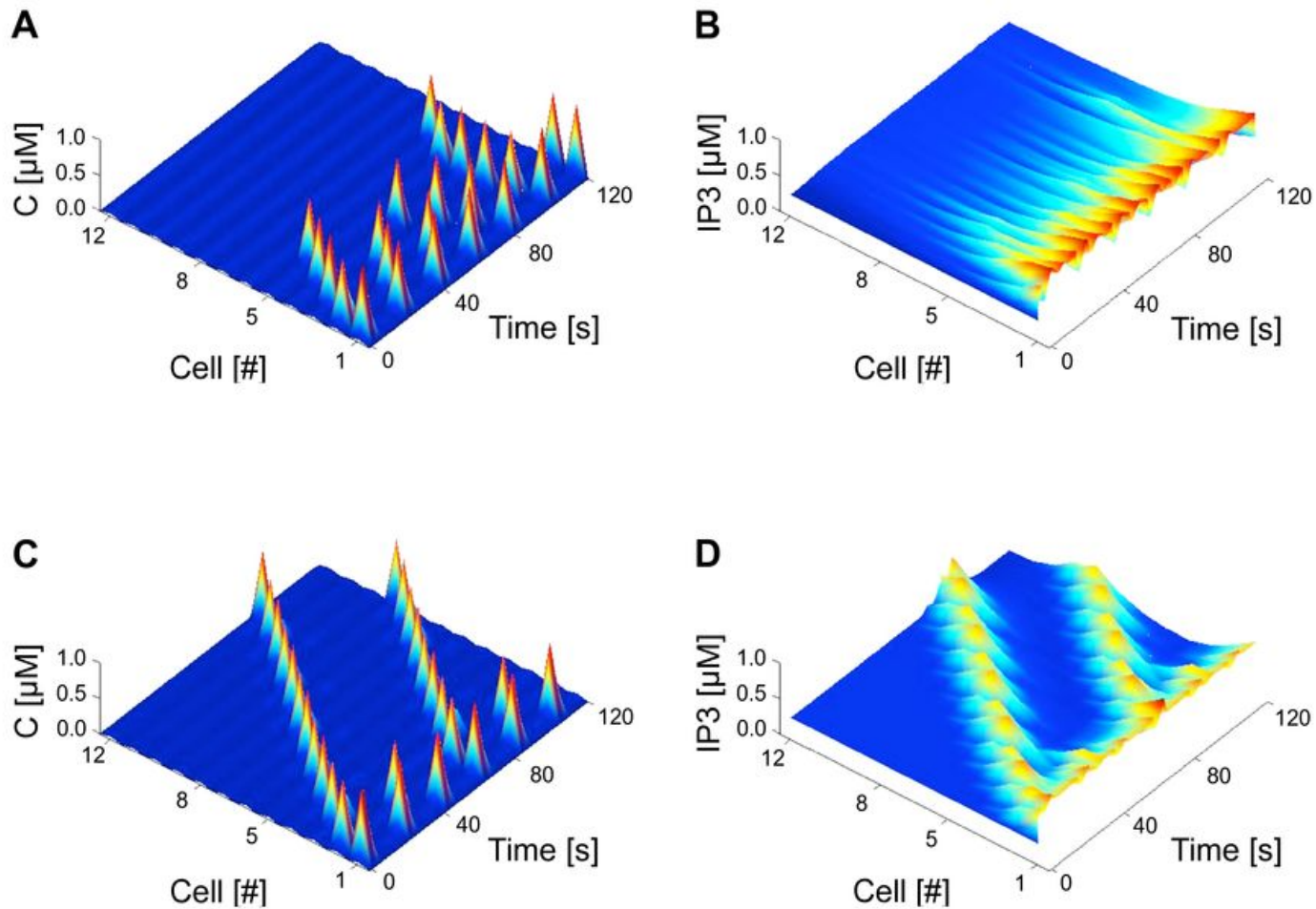

Figure 3

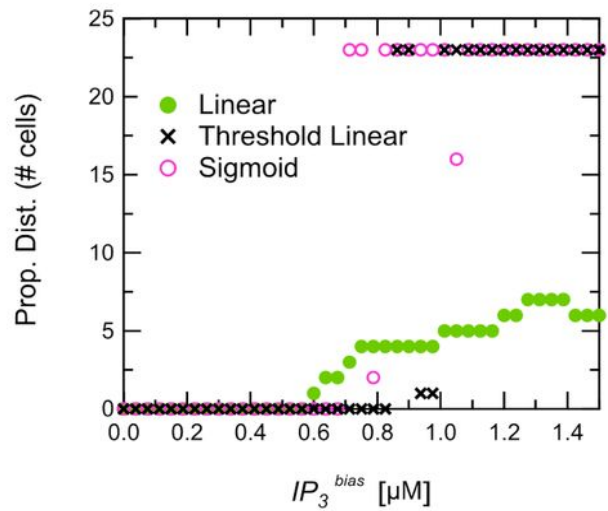 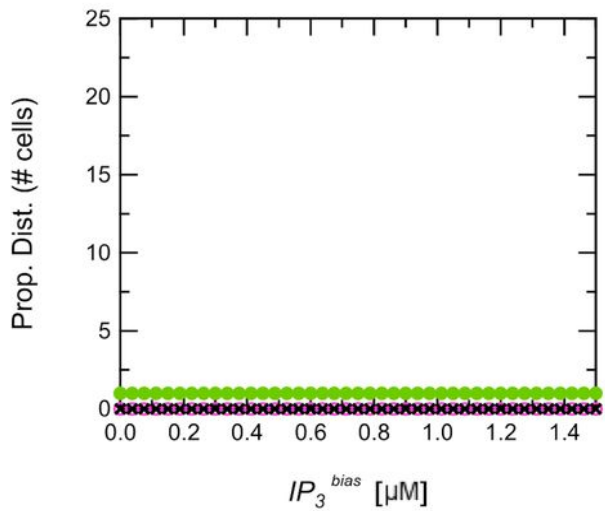
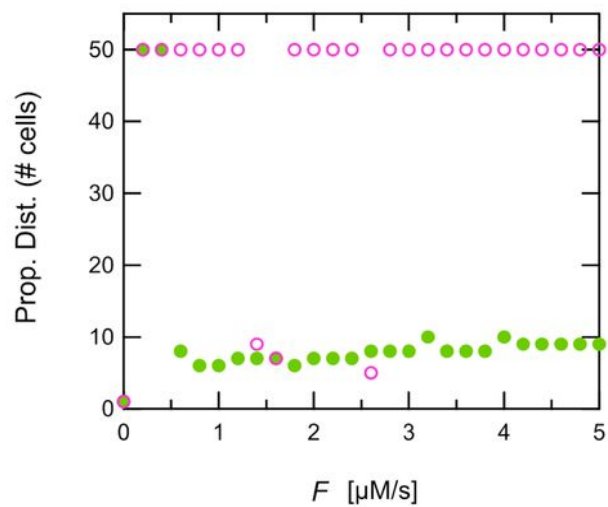 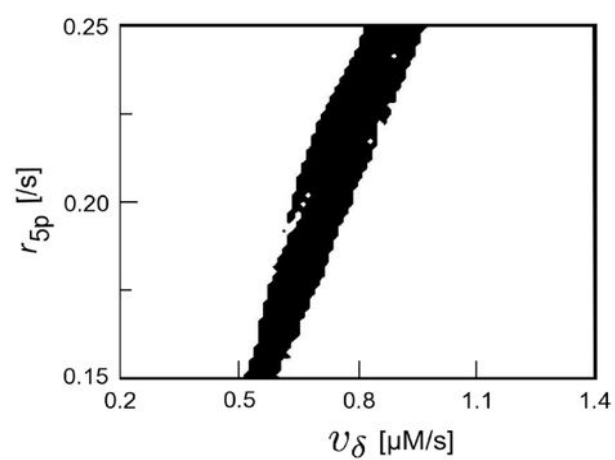

Figure 4

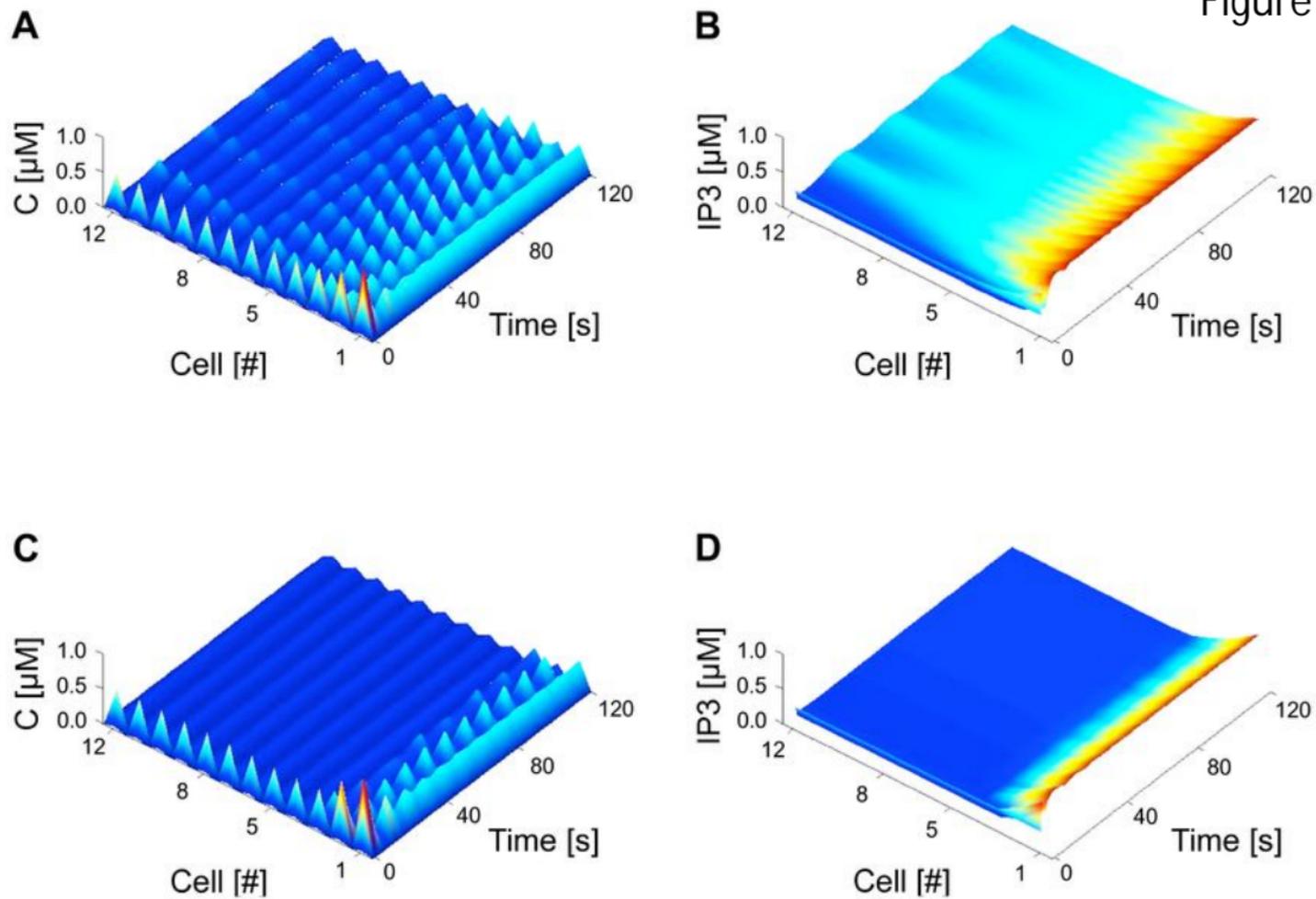

Figure 5

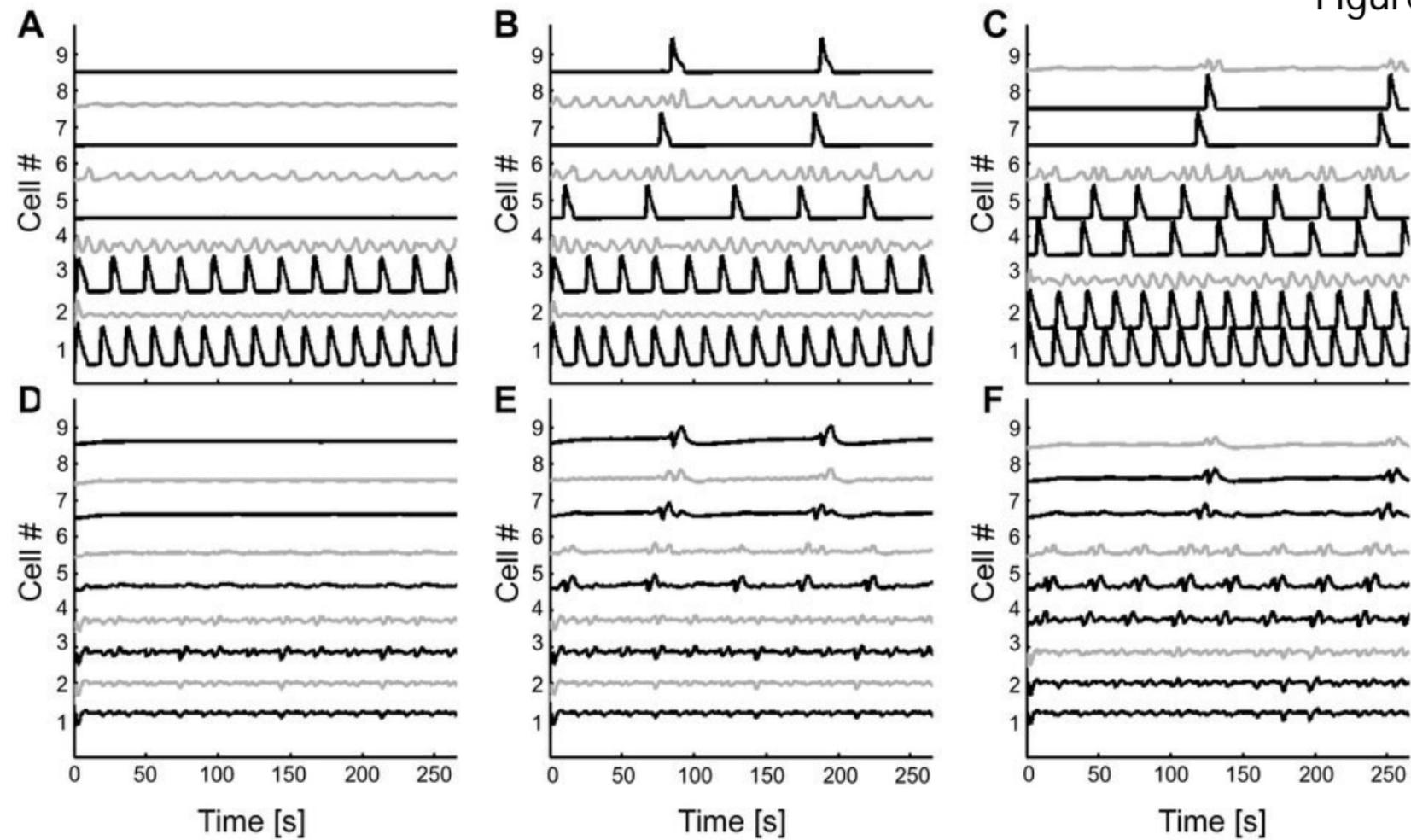

Figure 6

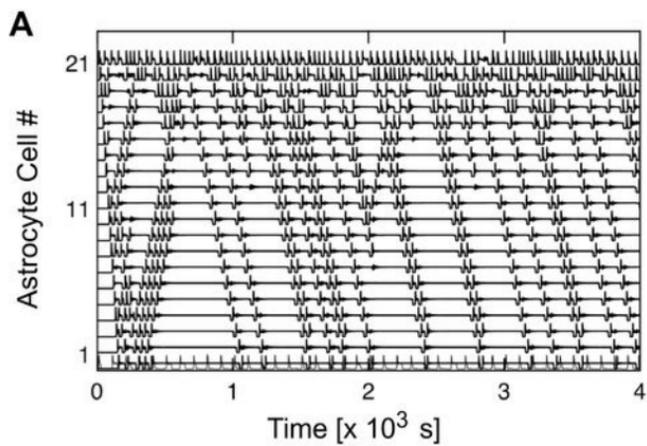
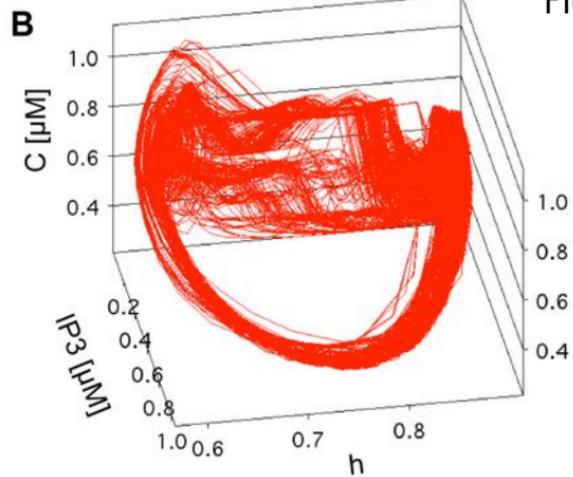
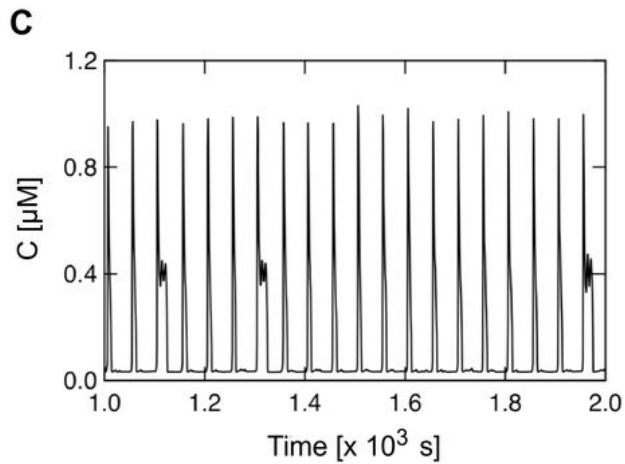
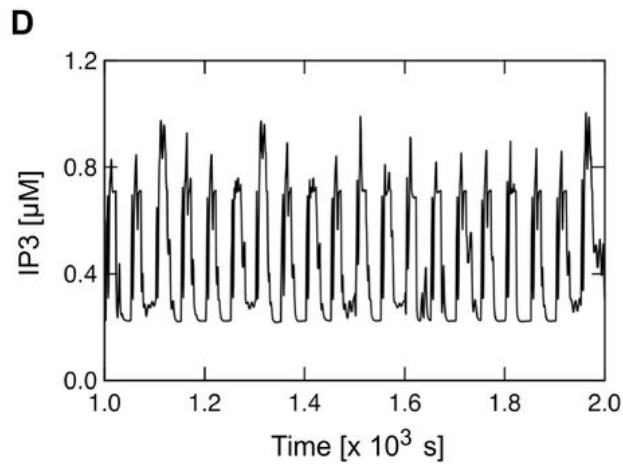

Figure 7

# Nonlinear gap junctions enable long-distance propagation of pulsating calcium waves in astrocyte networks

Mati Goldberg, Maurizio De Pittà, Vladislav Volman, Hugues Berry and Eshel Ben-Jacob

## SUPPLEMENTARY INFORMATION

## 1. The *ChI* model for intracellular $Ca^{2+}$ dynamics

Intracellular calcium dynamics in cells are the result of an intricate signaling pathway that mostly rests on $Ca^{2+}$ exchange between the cytoplasm and internal stores, such as the endoplasmic reticulum (ER) or mitochondria [1]. In astrocytes, $Ca^{2+}$ exchange with the ER plays the prominent role [2] and the concentration of cytoplasmic $Ca^{2+}$ depends on three distinct fluxes (illustrated in Figure 1a).

(1) Transfer of $Ca^{2+}$ from the cytoplasm to the ER lumen is mediated by an active pump, the (Sarco-)Endoplasmic-reticulum $Ca^{2+}$ ATPase (SERCA). SERCA contributes to the maintenance of higher $Ca^{2+}$ concentrations in the ER than in the cytoplasm.

(2) This concentration gradient drives a passive $Ca^{2+}$ leak from the ER to the cytoplasm.

(3) Specific $Ca^{2+}$ transporters, called inositol triphosphate receptor-channels (IP$_3$Rs, for short), contribute a third flux, transporting molecules from the ER to the cytoplasm. The regulation of IP$_3$Rs is remarkable: their activity is upregulated by a small molecule, inositol triphosphate (IP$_3$) and by small-to-intermediate cytoplasmic $Ca^{2+}$ levels. However, large $Ca^{2+}$ concentrations inhibit IP$_3$R activity.

Therefore, a small increase in cytoplasmic $Ca^{2+}$ levels activates IP$_3$Rs, which leads to increased $Ca^{2+}$ transport from the ER, ultimately increases cytoplasmic $Ca^{2+}$. Because of this positive feedback, $Ca^{2+}$ transport through IP$_3$Rs is usually called $Ca^{2+}$-induced $Ca^{2+}$ release (CICR). At some point however, the concentration of $Ca^{2+}$ within the cytoplasm becomes large enough to inhibit IP$_3$R activity and the surge of $Ca^{2+}$ through IP$_3$R terminates. These three transport mechanisms are classically captured by a mathematical model due to Li and Rinzel [3-5], given by eq.(1) and eq.(2) in the article. As is obvious from these equations, the Li-Rinzel (LR) model has two dynamical variables: $C$, the free cytosolic $Ca^{2+}$ concentration and $h$, the fraction of inactive (closed) IP$_3$R channels. The concentration of IP$_3$ in the cytoplasm (*IP$_3$*) instead is a parameter. To simulate this system, one must specify the mathematical formulation of the three fluxes $J_{pump}$, $J_{leak}$ and $J_{chan}$ as well as those of the steady fraction of closed IP$_3$Rs, $h_\infty$ and of the channel open/close transition time constant, $\tau_h$. In the present work, we have used the classical formulations originally introduced by Li and Rinzel [3-5].

As mentioned above, one feature of the Li-Rinzel model is that it treats the concentration of IP$_3$ as a mere parameter, i.e. *IP$_3$* in equations (1-2) is a constant value in time and does not change when the dynamical variables $C$ and $h$ vary over time. Yet, inspection of the intracellular $Ca^{2+}$ signaling pathway (Figure 1b) immediately shows that IP$_3$ concentration in the cytosol is coupled to $Ca^{2+}$ dynamics, and therefore is expected to change with $C$. In particular, IP$_3$ formation and degradation are both regulated by cytoplasmic $Ca^{2+}$.

We recently presented a model called the *ChI* model, that basically extends the Li-Rinzel model but includes IP$_3$ as a third dynamical variable [5]. To this aim, we had to simplify the complexity of the IP$_3$ signaling pathway (Figure 1c, d). This allowed us to derive a tractable yet realistic model. For clarity, we briefly describe the *ChI* model below. Further details and analyses can be found in [5].

In astrocytes, endogenous IP$_3$ dynamics (Figures 1c,d) mainly result from three enzymatic processes [6-8]. Endogeneous IP$_3$ is produced from a membrane lipid called phospatidylinositol biphosphate (PIP$_2$) in a reaction catalyzed by phospholipase Cδ (PLCδ). It can be degraded through phosphorylation (catalyzed by IP$_3$ 3-kinase, IP$_3$-3K) or through dephosphorylation (catalyzed by inositol polyphosphate 5-phosphatase, IP-5P). Experimentally, PLCδ is activated by cytoplasmic $Ca^{2+}$ in a cooperative fashion with a Hill exponent close to 2. But PLCδ is also inhibited by IP$_3$ concentrations larger than circa 1 μM. We thus expressed IP$_3$ production through PLCδ as:

$$P_{PLC\delta}(C, IP_3) = v_\delta \frac{K_I}{K_I + IP_3} \text{Hill}(C^2, K_{PLC\delta}) \tag{S1}$$

where $v_\delta$ is the maximal rate of IP$_3$ production by PLCδ, and we define generic Hill functions as $\text{Hill}(x^n, K) \equiv x^n/(x^n + K^n)$. Dephosphorylation of IP$_3$ by IP-5P displays pure Michaelis-Menten kinetics, but its Michaelis constant (≈10 μM) is higher than physiological IP$_3$ levels. It can thus be expressed as a linear reaction term:

$$D_{5P}(IP_3) = r_{5P} IP_3 \tag{S2}$$

As evidenced in Figure 1d, the regulation of IP$_3$ 3-kinase by cytoplasmic Ca$^{2+}$ is a complicated mechanism and includes interactions between Ca$^{2+}$, calmodulin and CaMKII. Using the algebraic properties of Hill functions however, we have been able to lump all this complexity into the product of a 4$^{th}$-order Hill function of Ca$^{2+}$ with a first-order Hill function of IP$_3$ (see [5] for further explanation):

$$D_{3K}(C, IP_3) = v_{3K} \text{Hill}(C^4, K_D) \text{Hill}(IP_3, K_3) \tag{S3}$$

Taken together, the IP$_3$ concentration in the *ChI* model is a third dynamical variable, whose dynamics are given by:

$$\frac{dIP_3}{dt} = P_{PLC\delta}(C, IP_3) - D_{3K}(C, IP_3) - D_{5P}(C, IP_3) + v_{bias} \tag{S4}$$

where $v_{bias}$ defines an external input of IP$_3$ coming e.g. from a neighboring astrocyte cell coupled through gap junction. Note that equation (S4) yields to equation (3) in the article when $v_{bias}$=0. Equations (1-2) and (S1-S4) together define the *ChI* model of intracellular Ca$^{2+}$ dynamics.

## 2. Bifurcation analysis of the *ChI* model and encoding modes

*a) Isolated cell*

We start our analysis with a description of the dynamics of the uncoupled case, i.e. the *ChI* model equations (1-2) and (S1-S4). Figure S1 shows bifurcation diagrams when the input bias of IP$_3$ varies in equation (S4). For either small or large biases, the IP$_3$ production term in equation (S4) is balanced by the two degradation terms. This gives rise to two stable steady-(time independent) states (black thick lines in Figures S1a, b). Further analysis of equations (1-2, S1-S4) shows that between the two steady-state regimes, spontaneous oscillations of the three variables (*C*, *h* and *IP$_3$*) arise. The evolution of the envelop of these oscillations (minimal and maximal values during the cycle) with $v_{bias}$ is shown in 3D plots as thin gray lines (Figures S1a, b) or in 2D projections as full thick lines (Figures S1c, d).

The general scenario indicated above (i.e. low steady state then oscillations then high steady state as $v_{bias}$ increases) is qualitatively conserved when the cell parameters vary in realistic ranges. However, a key observation for the present work is that, depending on the value of some cell physiological parameters (e.g. the affinity of SERCA pumps), the *nature* of the obtained oscillatory behavior changes. The major change concerns the type of bifurcation that is associated with the birth of the oscillatory behavior. Oscillations can arise through a Hopf bifurcation (Figure S1a) or through a Saddle-Node-on-Invariant-Circle (SNIC) bifurcation (Figures S1b). Now, the nature of the bifurcation decides how a change in the incoming IP$_3$ bias is encoded in the Ca$^{2+}$ oscillations. In fact one can classify these encoding behaviors by three classes.

(1) The amplitude of Ca$^{2+}$ oscillations more than doubles across the whole oscillatory range while their frequency stays almost constant. We refer to this behavior as the amplitude-modulation (AM) encoding mode.

(2) The frequency of Ca$^{2+}$ oscillations more than doubles across the whole oscillatory range while their amplitude stays almost constant. We refer to this behavior as the frequency-modulation (FM) encoding mode.

(3) Both the amplitude and the frequency of $Ca^{2+}$ oscillations change by more than a factor of 2 across the whole oscillatory range. We refer to this mode as the mixed amplitude-frequency-modulation (AFM) encoding mode.

Note that the waveform of the oscillations is also affected by the difference of the bifurcation type: FM-type oscillations usually display pulse-like waveforms, whereas AM-type ones are closer to sinusoidal [5].

In Figure S1 we illustrate the encoding modes displayed by the *ChI* model. The left-hand column (Figures S1a, c, e) illustrates a typical AFM behavior: increasing the $IP_3$ input bias yields an increase of the oscillation amplitude of $IP_3$ and *C* oscillations (Figure S1c) but also a marked decrease of their period (Figure S1e). The situation is more complex for the parameters used in the right-hand column: here the period varies dramatically but the amplitude of intracellular $Ca^{2+}$ oscillations is close to constant (though the amplitude of $IP_3$ oscillations varies). This corresponds to a FM regime (at least for *C*). Furthermore, as $Ca^{2+}$ transport through the SERCA pumps is larger for FM conditions [9], the range of $IP_3$ bias for which oscillations are predicted, is much narrower in the AFM regime than in the FM one.

*b) Coupled cell network*

We now consider a linear chain made of *N ChI* cells coupled to their two nearest neighbors by gap junctions as defined by equations (7-13). In the case of nonlinear coupling at least, one could expect that the dynamics of the cells would be different from the behavior predicted in the uncoupled (isolated) case studied above. To evaluate these changes, we first computed bifurcation diagrams for the coupled system (while Figure S1 shows bifurcation diagrams for isolated cells).

Unfortunately, deriving such bifurcation diagrams is arduous and uncertain from a numerical point of view because the resulting system is strongly nonlinear. As a consequence, the corresponding diagrams contain imperfectly resolved parts (see Figure S2 caption). Figure S2 shows bifurcation diagrams for $N = 7$ FM-encoding cells coupled by sigmoid-like gap junctions with reflective boundaries. We apply the $IP_3$ stimulus on the central cell in the chain (namely Cell 4 in Figure S2). In this case, we can consider only half of the chain and plot bifurcation diagrams for cells 4 to 7 only, since for symmetry reasons cells 3 to 1 display the same behavior as cells 5 to 7.

The bifurcation diagram of the central cell is qualitatively similar to the uncoupled case (compare Cell 4 in Figure S2 with Figures S1b, d). In particular, oscillations still arise through a SNIC bifurcation (at $IP_3^{bias} \approx 0.72$ µM). On the other hand, the bifurcation diagram beyond the SNIC displays multiple period-doubling cascades as well as different co-existing stable oscillatory regimes. As a consequence, the oscillatory regime in the central cell rapidly becomes extremely complex. The bifurcation diagrams for the other cells are also globally similar to that of the central (stimulated) cell, except that the fixed-point branches tend to get more horizontal when the distance from the stimulated cell increases. Furthermore, qualitatively similar diagrams are also obtained for different boundary conditions as well as for different stimulus locations (results not shown).

In general, these diagrams show FM pulse-like oscillatory regimes at low $IP_3^{bias}$ values, which can turn into complex oscillations for larger $IP_3^{bias}$ values. Yet because stable oscillation regimes can coexist in the bifurcation diagrams with stable fixed points, it cannot be predicted from these diagrams whether an $IP_3$ input arriving to a given cell from its neighbor will trigger pulse-like oscillations or not, i.e. whether it will switch the system from the fixed point to the oscillatory regime. Accordingly, we used extensive numerical simulations to investigate under what conditions one could observe intercellular propagation of $Ca^{2+}$ waves along the astrocyte chain.

In agreement with previous studies (see [10] for a review), $IP_3$-triggered CICR indeed allows intercellular $Ca^{2+}$ wave propagation in our modeling framework, as shown in Figure S3. Here the simple case of $N = 5$ FM *ChI* astrocytes coupled through sigmoid gap junctions is considered and the first cell (*cell A1*) is stimulated from $t = 10$ s to $t = 30$ s by a constant bias of $IP_3^{bias} = 0.8$ µM. In agreement with the bifurcation diagrams shown in Figure S2, pulse-like $Ca^{2+}$ oscillations are observed. Note that the 20-second long stimulation used here is shorter than the predicted period of oscillations at such $IP_3^{bias}$ value. Therefore, only a single pulse is observed. In general when a large-enough $IP_3$ pulse arrives to a given cell, either by direct external stimulation or from a neighbor coupled cell, CICR first generates a neat pulse-like $Ca^{2+}$ response (Figure S3, *black curves*), which predates a pulse-like increase of $IP_3$ (Figure S3, *dashed blue curves*). Part of this $IP_3$ then diffuses to the neighbor cells in the form of a new pulse. Therefore, as long as the newly formed $IP_3$ pulses are large enough, $IP_3$-triggered CICR is preserved and a propagating pulsating $Ca^{2+}$ wave, mediated by $IP_3$ diffusion, is observed.

On the contrary, lack of $IP_3$ stops wave propagation, as it happens for example in Figure S3 for the $IP_3$ pulse that leaves cell A3 for cell A4. This pulse is indeed too small to trigger CICR in cell A4, so that $Ca^{2+}$ wave propagation ceases beyond cell A3. Notably, the distance (in terms of number of cells) that a $Ca^{2+}$ wave can travel along the chain strongly depends both on the nature of the gap-junction coupling and on cell type (AFM or FM) as is shown in the article.

| Parameter | Value (AFM/FM) | Units | Description |
|---|---|---|---|
| *Li-Rinzel core parameters* | | | |
| $C_0$ | 2.0 | µM | Total cell free [$Ca^{2+}$] per cytosolic volume |
| $c_1$ | 0.185 | – | Ratio between ER and cytosol volumes |
| $r_C$ | 6.0 | $s^{-1}$ | Maximal CICR rate |
| $r_L$ | 0.11 | $s^{-1}$ | Maximal rate of $Ca^{2+}$ leak from the ER |
| $v_{ER}$ | 0.9 | $µM \cdot s^{-1}$ | Maximal SERCA uptake rate |
| $d_1$ | 0.13 | µM | $IP_3$ dissociation constant |
| $d_2$ | 1.049 | µM | $Ca^{2+}$ inactivation dissociation constant |
| $d_3$ | 0.9434 | µM | $IP_3$ dissociation constant |
| $d_5$ | 0.08234 | µM | $Ca^{2+}$ activation dissociation constant |
| $a_2$ | 0.200 | $s^{-1}$ | $IP_3R$ binding rate constant for $Ca^{2+}$ inhibition |
| $K_{ER}$ | 0.10 / 0.05 | µM | SERCA $Ca^{2+}$ affinity |
| *$IP_3$ metabolism parameters* | | | |
| $v_\delta$ | 0.12 / 0.7 | $µM \cdot s^{-1}$ | Maximal rate of $IP_3$ synthesis by PLCδ |
| $K_{PLC\delta}$ | 0.1 | µM | $Ca^{2+}$ affinity of PLCδ |
| $\kappa_\delta$ | 1.5 | µM | Inhibition constant of PLCδ activity |
| $v_{3K}$ | 4.5 | $s^{-1}$ | Rate of $IP_3$ degradation by 3K |
| $K_{3K}$ | 0.7 | µM | Half maximal degradation rate of $IP_3$ by $IP_3$-3K |
| $K_3$ | 1 | µM | Half-saturation constant for $Ca^{2+}$-dependent $IP_3$-3K activation |
| $r_{5P}$ | 0.04 / 0.21 | $s^{-1}$ | Rate of $IP_3$ degradation by IP-5P |
| $F$ | 2 | $µM \cdot s^{-1}$ | Gap junction permeability (or coupling strength) |
| *Nonlinear gap-junction parameters* | | | |
| $IP_3^{shift}$ | 0.3 | µM | Half-maximal diffusion $IP_3$ threshold |
| $IP_3^{scale}$ | 0.05 | µM | Slope factor |
| $IP_3^{bias}$ | 0.8 / 1.0 | µM | Imposed $IP_3$ level in the stimulated cell |

# Nonlinear gap junctions enable long-distance propagation of pulsating calcium waves in astrocyte networks

Mati Goldberg, Maurizio De Pittà, Vladislav Volman, Hugues Berry and Eshel Ben-Jacob

**SUPPLEMENTARY INFORMATION**

## Supplementary Figures - Captions

**Figure SI1**. Bifurcation analysis of an uncoupled (i.e. isolated) *ChI* astrocyte for AFM (**a**, **c**, **e**) and FM (**b**, **d**, **f**) encoding regimes. (**a**, **b**) 3D-rendering of bifurcation surfaces in the state space. AFM oscillation amplitude (**c**) and period (**e**) are controlled by a supercritical Hopf (H) bifurcation and a saddle-node limit cycle (SNC) bifurcation respectively. Conversely in FM-mode, the occurrence of a saddle-node on an invariant circle (SNIC) bifurcation accounts for the rise of arbitrarily-small frequency $Ca^{2+}$ oscillations (**f**) at almost constant amplitude (**d**). Legend: (**a**, **b**): *black lines*: stable fixed points; *red dashed lines* unstable fixed points; *blue lines*: bifurcating limit cycles; semi-transparent *surfaces* denote envelopes of stable (*grey*) and unstable (*red*) oscillations. (**c-f**): *green*: $IP_3$; *orange*: $Ca^{2+}$ (**c**) *full lines*: stable oscillations; *dashed lines*: unstable oscillations. Parameters as in Table 1.

**Figure SI2**. Bifurcation analysis of the astrocyte chain model for $N = 7$ FM-encoding cells with sigmoid coupling and reflective boundary conditions. Calcium concentrations at steady states are shown for the central (stimulated) cell (*Cell 4*) and for cells 5, 6 and 7. Although not apparent in the figure, for $IP_3^{bias}$ values larger than $\approx 0.8$ μM, the stable oscillations become far more complex than in the isolated case. This is due to a very rapid cascade of period-doubling bifurcations, which yields extremely complex limit cycles (with numerous folds) that could not be precisely rendered in the figure (see also Section III.1.b). Moreover, for $IP_3^{bias} > 1.1$ μM, the amplitude of these limit cycles shrinks and numerical investigations evidenced the coexistence of multiple complex stable orbits. Legend: *thin full lines* locate unstable fixed points, and *thick full*

*lines* stable one. *Full* (*open*) *circles* denote the envelopes of *stable* (*unstable*) limit cycles. Letters denote bifurcation type as in Figure SI1.

**Figure SI3**. IP$_3$-triggered CICR-mediated propagation of a pulsed Ca$^{2+}$ wave within a chain of five FM *ChI* astrocytes (A1-A5). (**a**) An IP$_3$ stimulation of constant intensity ($IP_3^{bias}$ = 0.8 µM) is applied to cell A1 from $t$ = 10 s to $t$ = 30 s. This increases IP$_3$ concentration, thus triggering CICR from the ER and the generation of a Ca$^{2+}$ pulse. (**b**) By means of communication through gap junctions, suprathreshold IP$_3$ from A1 can diffuse to A2, triggering CICR there. The process is essentially regenerative so that a Ca$^{2+}$ pulse almost identical to the original one can be observed in the arrival cells. (**c**) As soon as the IP$_3$ influx to one cell from its neighbors is not sufficient to trigger CICR, the propagation stops. This is indeed the case of cells A4 and A5. Cells were coupled by sigmoid gap junctions and experienced reflective boundary conditions.

**Figure SI4**. Propagation patterns with non-linear sigmoid-like gap junctions in an astrocyte chain of 12 FM-encoding cells with periodic (**a,b**) or absorbing (**c,d**) boundary conditions. Stimulation triggered by $IP_3^{bias}$ = 1.0 µM from $t$ = 0 s to $t$ = 120 s applied to the central cell (i.e. cell 6) (**a,b**) or the first cell in the line (**c,d**).

**Figure SI5**. Calcium traces for wave propagation in composite astrocyte chains constituted of both FM (*black*) and AFM (*gray*) cells with (**a**) $IP_3^{thr}$ = 0.215 µM or (**b**) $IP_3^{thr}$ = 0.3 µM. The diffusion threshold is critical to determine the efficiency of transmission of Ca$^{2+}$ waves along astrocyte chains. Other parameters as in Figure 7c.

**Figure SI6**. Interpulse interval distributions for the simulations shown in Figure 10. (**a**) Distributions for cells A1 to A10. (**b**) Distributions for cells A11 to A20. In several cells (e.g. cells A1, A2, A4 and A5), the distribution is broad, and large intervals are as commonly observed as smaller ones. Each panel indicate the number of pulses counted.

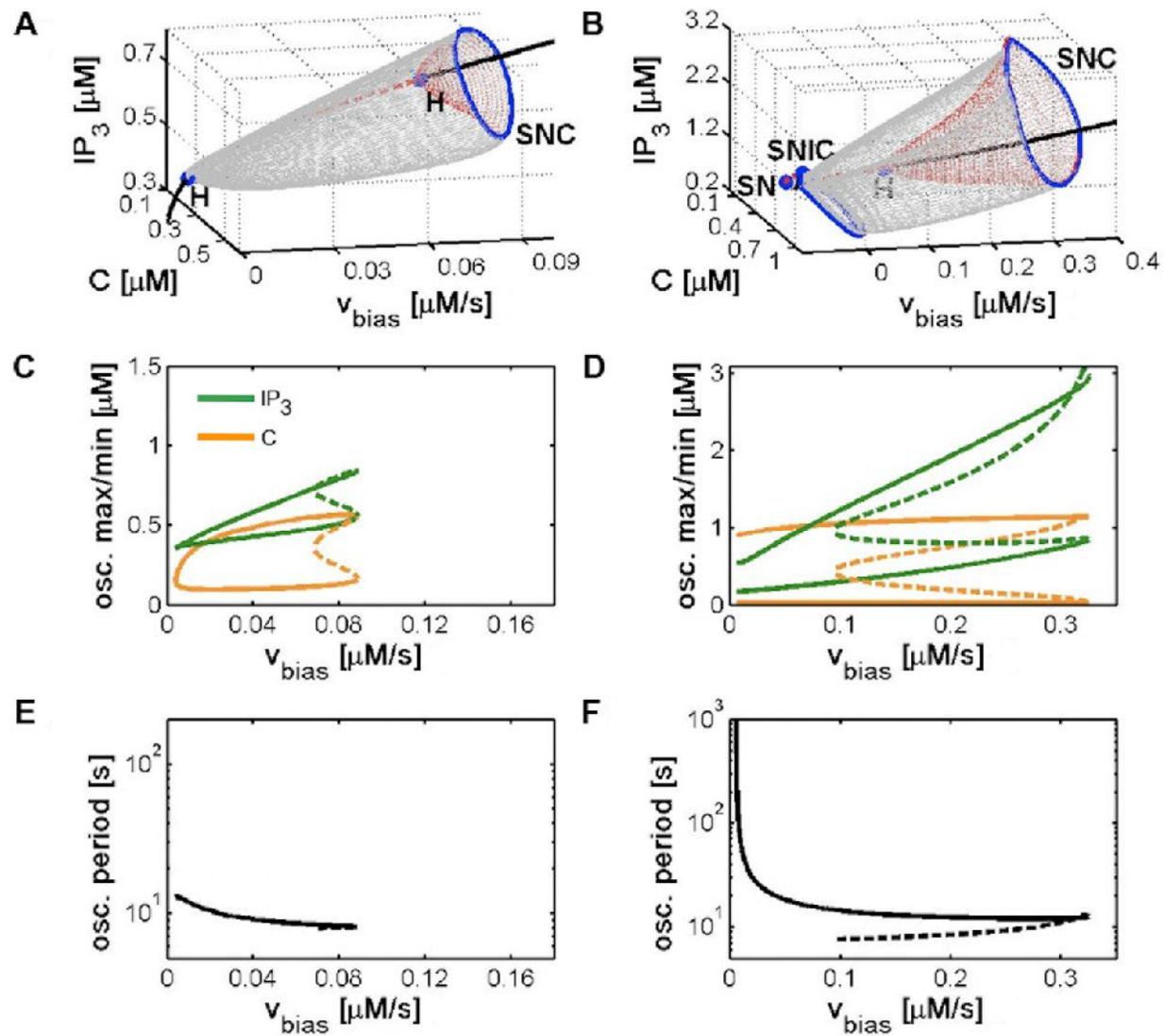

Supplementary Figure 1

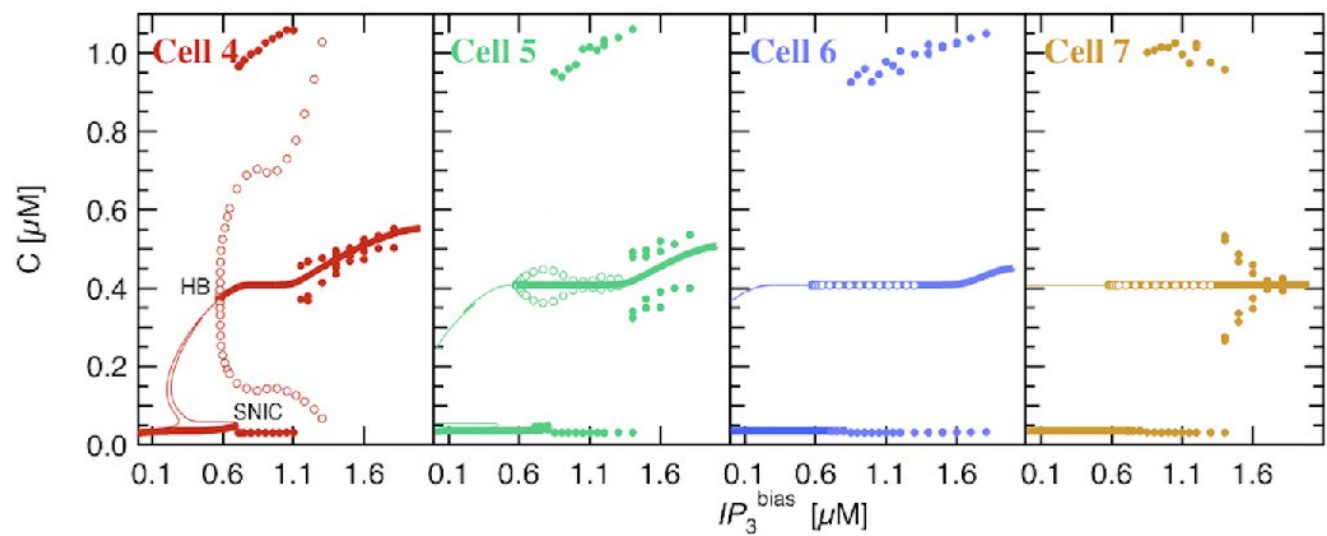

Supplementary Figure 2

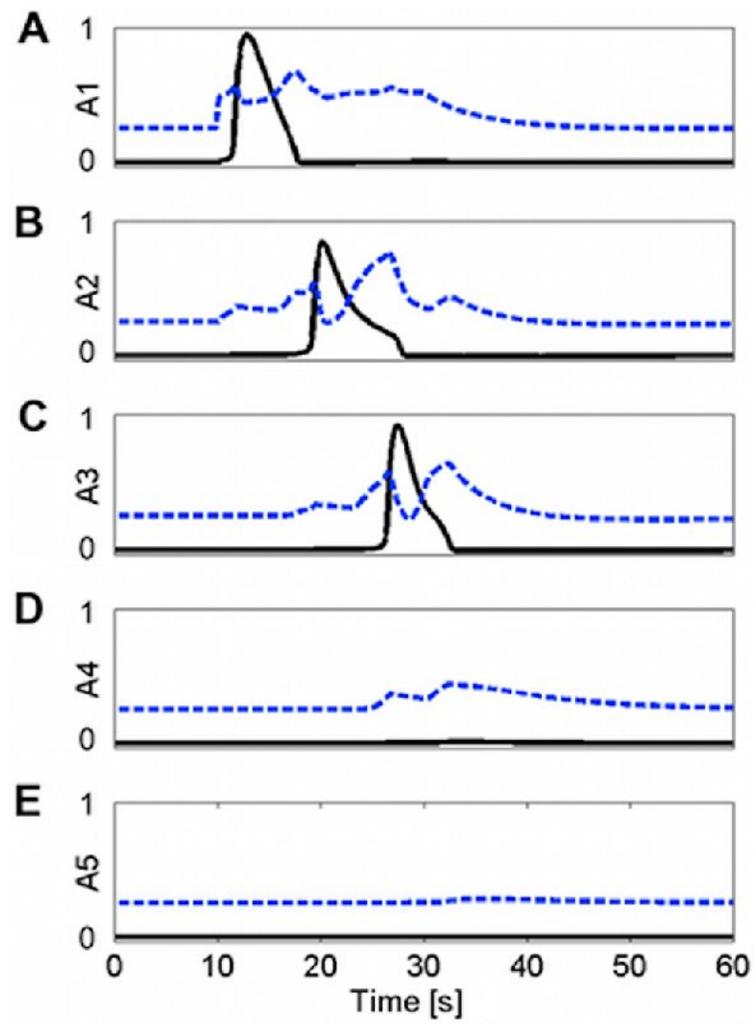

Supplementary Figure 3

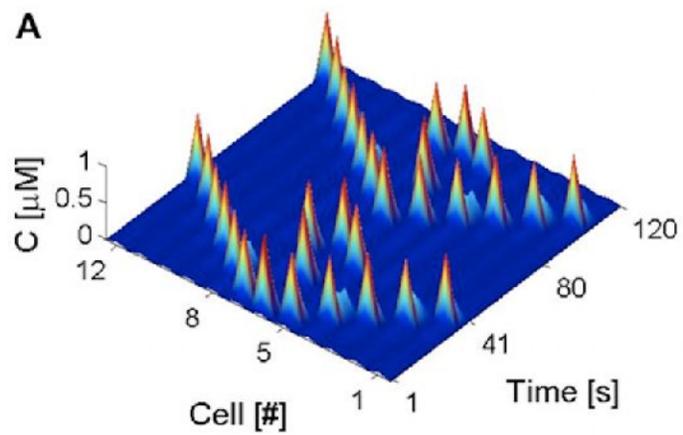
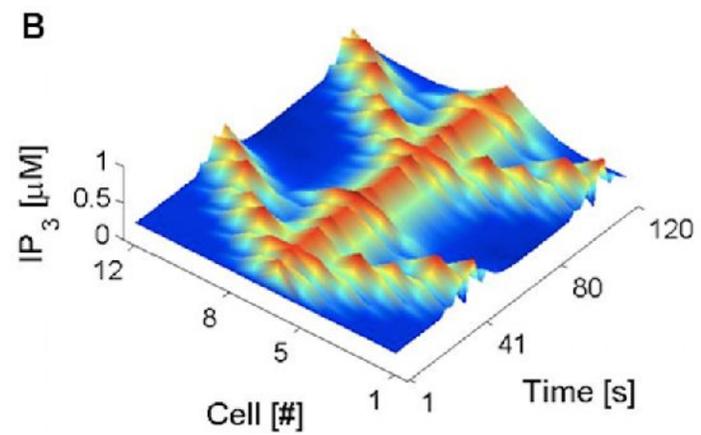
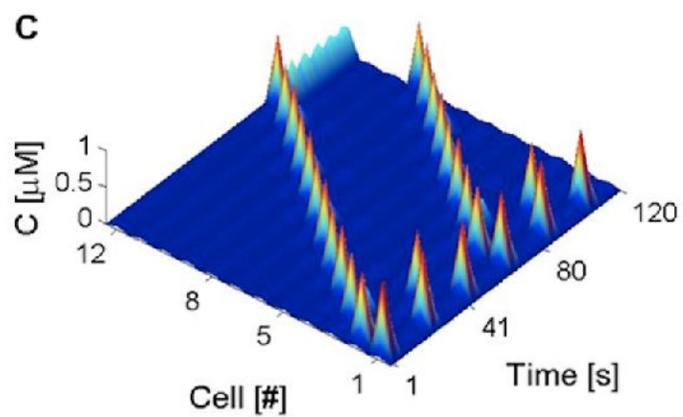
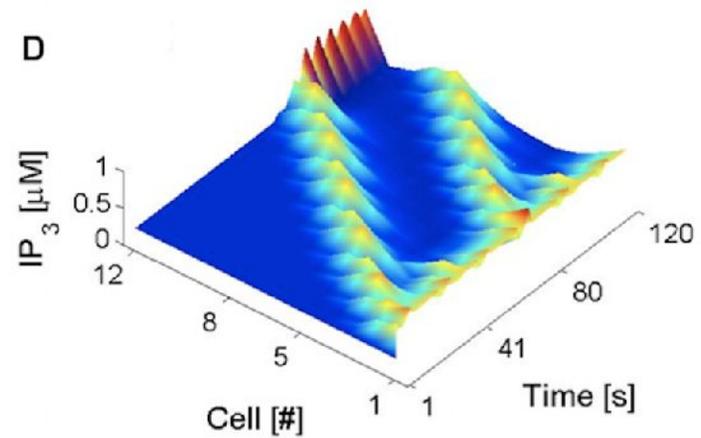

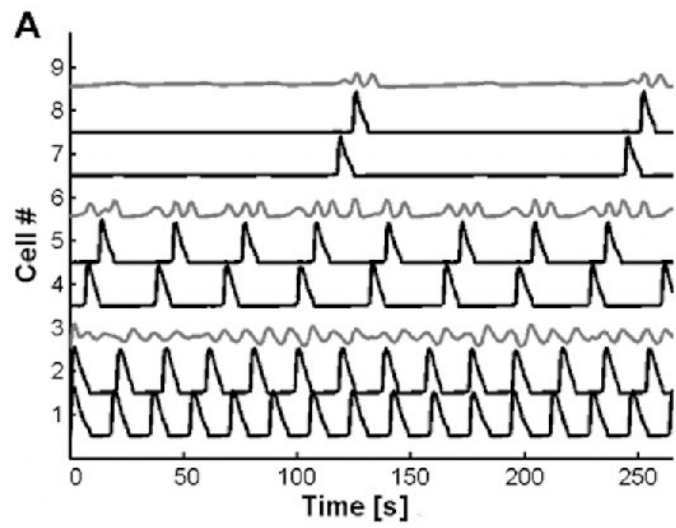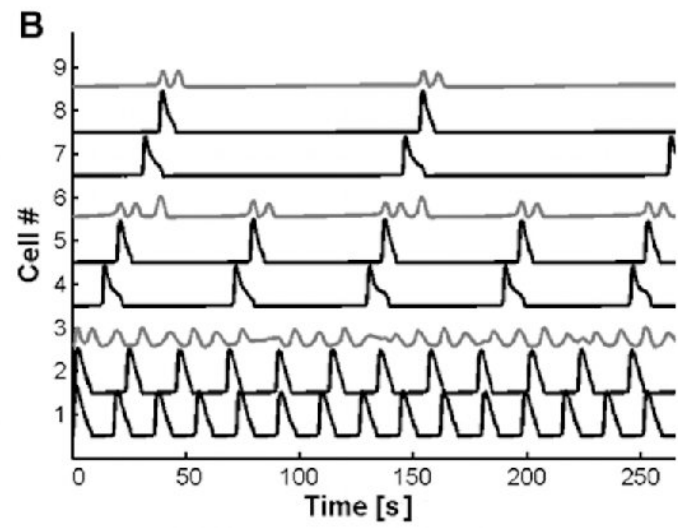

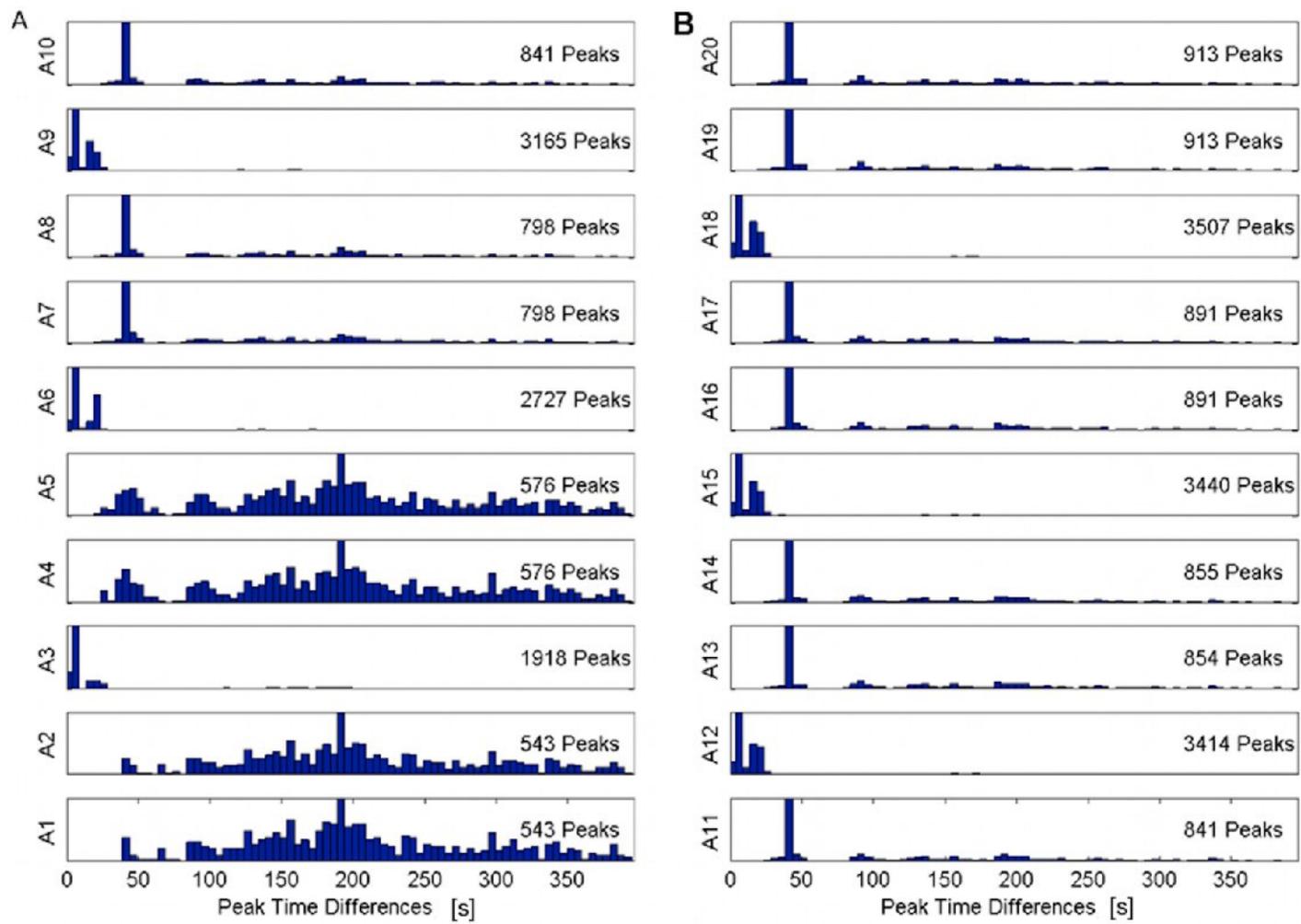

Supplementary Figure